\documentclass[11pt,a4paper]{article}
\usepackage{jcappub}
\usepackage{lmodern}
\usepackage{graphicx,color}
\graphicspath{{figures/}}
\usepackage{comment}
\usepackage{cancel}

\usepackage[T1]{fontenc}
\usepackage[utf8]{inputenc}
\usepackage{color}
\usepackage{amsmath}
\usepackage{amssymb}
\usepackage{esint}
\usepackage{appendix}
\usepackage{bm}
\usepackage{url}
\usepackage{hyperref}
\usepackage{xcolor}
\usepackage{mathtools} 
\newif\ifdraft
\draftfalse 

\usepackage[normalem]{ulem}
\begin{document}

\makeatletter


\definecolor{somegreen}{cmyk}{0,0.49,0.98,0.09}
\definecolor{red}{rgb}{1,0,0}
\definecolor{magenta}{cmyk}{0,1,0,0}
\definecolor{lavender}{cmyk}{0.16,0.67,0,0.57}
\definecolor{darkgreen}{rgb}{0,0.65,0.05}
\definecolor{antiquefuchsia}{rgb}{0.33, 0.1, 0.89}

\newcommand{\fontsmall}{\fontsize{8pt}{9pt}\selectfont}
\def\Mp{M_{\ast}}
\def\beqra{\begin{eqnarray}}
\def\eeqra{\end{eqnarray}}
\def\beq{\begin{equation}}
\def\eeq{\end{equation}}
\def\etap{\eta^\prime}
\def\etain{\eta_{in}}
\def\ds{\displaystyle}
\def\ts{\textstyle}
\def\ss{\scriptstyle}
\def\sss{\scriptscriptstyle}
\def\Vb{\bar{V}}
\def\vp{\bar{\varphi}}
\def\chb{\bar{\chi}}
\def\ab{\bar{a}}
\def\rhb{\bar{\rho}}
\def\Phib{\bar{\Phi}}
\def\Psib{\bar{\Psi}}
\def\L{\Lambda}
\def\l{\lambda}
\def\kp{\kappa}
\def\vp{\varphi}
\def\T{\Theta}
\def\qm{q_{\mathrm{min}}}
\def\Qm{Q_{\mathrm{min}}}
\def\sm{\sigma_{\mathrm{min}}}
\def\bx{{\bf{x}}}
\def\bk{{\bf{k}}}
\def\bp{{\bf{p}}}
\def\bq{{\bf{q}}}
\def\bu{{\bf{u}}}
\def\bv{{\bf{v}}}
\def\bn{{\bf{n}}}
\def\bm{{\bf{m}}}
\def\bl{{\bf{l}}}
\def\bw{{\bf{w}}}
\def\bl{{\bf{l}}}
\def\bL{{\bf{L}}}
\def\iv{{\bf i_{\rm v}}}
\def\jv{{\bf j_{\rm v}}}
\def\kv{{\bf k_{\rm v}}}
\def\rv{{\bf r_{\rm v}}}
\def\thetav{{\mathbf{\theta}_{\rm v}}}
\def\phiv{{\bf \phi_{\rm v}}}
\def\bV0{{\bf{V_0}}}
\def\re#1{(\ref{#1})}
\def\D{\Delta}
\def\G{\Gamma}
\def\gmn{g_{\mu\nu}}
\def\tgmn{ \tilde{g}_{\mu\nu}}
\def\p{\partial}
\def\half{\mbox{\small$\frac{1}{2}$}}
\def\de{\delta}
\def\rM{\rho_\chi}
\def\rr{\rho_{\mathrm rad}}
\def\rb{\rho_{\mathrm b}}
\def\rvp{\rho_\vp}
\def\agt{~\mbox{\raisebox{-.6ex}{$\stackrel{>}{\sim}$}}~}
\def\alt{~\mbox{\raisebox{-.6ex}{$\stackrel{<}{\sim}$}}~}
\def\bx{{\bf{x}}}
\def\by{{\bf{y}}}
\def\br{{\bf{r}}}
\def\bl{{\bf{l}}}
\def\bs{{\bf{s}}}
\def\bk{{\bf{k}}}
\def\bp{{\bf{p}}}
\def\bq{{\bf{q}}}
\def\bu{{\bf{u}}}
\def\bv{{\bf{v}}}
\def\bV{{\bf{V}}}
\def\bw{{\bf{w}}}
\def\bl{{\bf{l}}}
\def\aa{\`a }
\def\ee{\`e }
\def\eea{\'e  }
\def\oo{\`o }
\def\uu{\`u }
\def\ii{\`i }
\def\bz{{\bf{z}}}
\def\Wt{\tilde W}

\def\vpb{\bar{\vp}}
\def\vp{\varphi}

\def\Ng{N_{\rm g}} 

\def\be{\begin{equation}}
\def\ee{\end{equation}}
\newcommand{\vs}{\nonumber\\}
\def\ba#1\ea{\begin{align}#1\end{align}}
\def\refeq#1{Eq.~\eqref{eq:#1}}
\def\v#1{\mathbf{#1}}

\newcommand{\avg}[1]{\left\langle #1 \right\rangle}
\newcommand{\WtwoProd}[1]{\prod_{r=1}^{#1} W_2(n_r)}

\newcommand{\diff}{\mathop{}\!\mathrm{d}}
\newcommand{\hMpc}{h\,\mathrm{Mpc}^{-1}}
\newcommand{\kmin}{k_\mathrm{min}}
\newcommand{\kmax}{k_\mathrm{max}}
\newcommand{\Ngrid}{N_\mathrm{grid}}

\newcommand{\Deltaj}[2]{%
\Big(\mathcal{H}_{#1+2}(#2,-#2)-\mathcal{H}_{#1}\mathcal{H}_2(#2,-#2)\Big)}
\newcommand{\prodn}[1]{\prod_{r=1}^{#1} W_2(n_r)}

\newcommand{\fNL}{f_\mathrm{NL}}
\newcommand{\fNLloc}{f_\mathrm{NL}^\mathrm{local}}
\newcommand{\fNLeq}{f_\mathrm{NL}^\mathrm{equil}}
\newcommand{\fNLort}{f_\mathrm{NL}^\mathrm{ortho}}
\newcommand{\Mmin}{M_\mathrm{min}}
\newcommand{\sump}{\mathop{\sum\nolimits^{\prime}}}

\newcommand{\Planck}{\textit{Planck}}
\newcommand{\Quijote}{\textsc{Quijote}}
\newcommand{\QuijotePNG}{\textsc{Quijote-png}}

\def\la{~\mbox{\raisebox{-.6ex}{$\stackrel{<}{\sim}$}}~}
\def\ga{~\mbox{\raisebox{-.6ex}{$\stackrel{>}{\sim}$}}~}

\newcommand{\mq}{\textcolor{magenta}}
\newcommand{\mqnote}[1]{\textcolor{red}{[{\bf MQ}: #1]}}
\newcommand{\rev}{\textcolor{blue}}
\newcommand{\dd}{\textrm{d}}

\newcommand{\MP}[1]{{\textcolor{maroon}{\bf #1}}\marginpar{\textcolor{maroon}{$\bullet$ MP}}}

\newcommand{\GB}[1]{{\textcolor{cyan}{\bf #1}}\marginpar{\textcolor{cyan}{$\bullet$ Giorgia}}}

\def\chg#1{\textcolor{magenta}{#1}} 

\let\jnl@style=\rm
\def\ref@jnl#1{{\jnl@style#1}}

\def\aj{\ref@jnl{AJ}}                   
\def\actaa{\ref@jnl{Acta Astron.}}      
\def\araa{\ref@jnl{ARA\&A}}             
\def\apj{\ref@jnl{ApJ}}                 
\def\apjl{\ref@jnl{ApJ}}                
\def\apjs{\ref@jnl{ApJS}}               
\def\ao{\ref@jnl{Appl.~Opt.}}           
\def\apss{\ref@jnl{Ap\&SS}}             
\def\aap{\ref@jnl{A\&A}}                
\def\aapr{\ref@jnl{A\&A~Rev.}}          
\def\aaps{\ref@jnl{A\&AS}}              
\def\azh{\ref@jnl{AZh}}                 
\def\baas{\ref@jnl{BAAS}}               
\def\bac{\ref@jnl{Bull. astr. Inst. Czechosl.}}
\def\caa{\ref@jnl{Chinese Astron. Astrophys.}}
\def\cjaa{\ref@jnl{Chinese J. Astron. Astrophys.}}
\def\icarus{\ref@jnl{Icarus}}           
\def\jcap{\ref@jnl{J. Cosmology Astropart. Phys.}}
\def\jrasc{\ref@jnl{JRASC}}             
\def\memras{\ref@jnl{MmRAS}}            
\def\mnras{\ref@jnl{MNRAS}}             
\def\na{\ref@jnl{New A}}                
\def\nar{\ref@jnl{New A Rev.}}          
\def\pra{\ref@jnl{Phys.~Rev.~A}}        
\def\prb{\ref@jnl{Phys.~Rev.~B}}        
\def\prc{\ref@jnl{Phys.~Rev.~C}}        
\def\prd{\ref@jnl{Phys.~Rev.~D}}        
\def\pre{\ref@jnl{Phys.~Rev.~E}}        
\def\prl{\ref@jnl{Phys.~Rev.~Lett.}}    
\def\pasa{\ref@jnl{PASA}}               
\def\pasp{\ref@jnl{PASP}}               
\def\pasj{\ref@jnl{PASJ}}               
\def\rmxaa{\ref@jnl{Rev. Mexicana Astron. Astrofis.}}%
\def\qjras{\ref@jnl{QJRAS}}             
\def\skytel{\ref@jnl{S\&T}}             
\def\solphys{\ref@jnl{Sol.~Phys.}}      
\def\sovast{\ref@jnl{Soviet~Ast.}}      
\def\ssr{\ref@jnl{Space~Sci.~Rev.}}     
\def\zap{\ref@jnl{ZAp}}                 
\def\nat{\ref@jnl{Nature}}              
\def\iaucirc{\ref@jnl{IAU~Circ.}}       
\def\aplett{\ref@jnl{Astrophys.~Lett.}} 
\def\apspr{\ref@jnl{Astrophys.~Space~Phys.~Res.}}
\def\bain{\ref@jnl{Bull.~Astron.~Inst.~Netherlands}}
\def\fcp{\ref@jnl{Fund.~Cosmic~Phys.}}  
\def\gca{\ref@jnl{Geochim.~Cosmochim.~Acta}}   
\def\grl{\ref@jnl{Geophys.~Res.~Lett.}} 
\def\jcp{\ref@jnl{J.~Chem.~Phys.}}      
\def\jgr{\ref@jnl{J.~Geophys.~Res.}}    
\def\jqsrt{\ref@jnl{J.~Quant.~Spec.~Radiat.~Transf.}}
\def\memsai{\ref@jnl{Mem.~Soc.~Astron.~Italiana}}
\def\nphysa{\ref@jnl{Nucl.~Phys.~A}}   
\def\physrep{\ref@jnl{Phys.~Rep.}}   
\def\physscr{\ref@jnl{Phys.~Scr}}   
\def\planss{\ref@jnl{Planet.~Space~Sci.}}   
\def\procspie{\ref@jnl{Proc.~SPIE}}   

\makeatother
\title{On the Relation Between Field-Level Posteriors, Correlators, and their Likelihoods}

\author[a]{Massimo Pietroni}
\affiliation[a]{Dipartimento di Scienze Matematiche, Fisiche e Informatiche, Universit\`a di Parma, and INFN Gruppo Collegato di Parma, Parco Area delle Scienze 7/A, I-43124, Parma, Italy}
\author[b]{Fabian Schmidt}
\affiliation[b]{Max-Planck-Institut für Astrophysik, Karl-Schwarzschild-Str.\ 1, 85748 Garching, Germany}

\abstract{
  We develop a field-level posterior for cosmological data by marginalizing over initial conditions and noise in a general forward model.
While our focus is on large-scale structure data, the results generalize to any weakly non-Gaussian observable. Moreover,
  the construction is non-perturbative with respect to the forward model and applies equally well to perturbative calculations, simulation-based predictions, and more general effective descriptions. Expanding the FLP around its Gaussian limit, we derive a general expression for the Fisher matrix and reorganize the field-level information into contributions associated with the connected correlators of the evolved field. This makes explicit which terms are captured by likelihood analyses based on the power spectrum, the bispectrum, or finite sets of summary statistics, and which are lost under compression. We recover the standard Gaussian-covariance result for the power spectrum, show that the Gaussian bispectrum likelihood reproduces the corresponding field-level contribution, and show how cross-covariances among summaries progressively reconstruct more of the full field-level information. As an application to the BAO scale, we show how the field contains all the information required for its optimal reconstruction in the presence of noise, and identify the contributions in the FLP needed to attain this limit. We also show that the reconstruction of the initial field arises naturally as a byproduct of our approach, yielding the optimal estimate of the initial conditions given the data and the noise. Our results provide a unified framework to compare field-level and correlator-based inference, to quantify the information loss induced by compression, and to explore the role of stochasticity.
}

\emailAdd{\\massimo.pietroni@unipr.it\\fabians@MPA-Garching.mpg.de} 


\maketitle
\section{Introduction}
The statistical analysis of large-scale structure is often formulated in terms of summary statistics, most prominently the power spectrum and the bispectrum. This strategy is attractive both conceptually and computationally: it organizes the information into a small set of observables and allows one to build likelihoods directly for the corresponding estimators. At the same time, any such compression raises a basic question: which part of the information contained in the full field is retained, and which part is discarded? Answering this question systematically is becoming increasingly important as galaxy surveys move deeper into the mildly nonlinear regime and as forward models become more accurate and versatile.

Field-level inference offers a natural framework in which to pose this question. Instead of compressing the data at the outset, one writes a probabilistic description directly for the observed field, given a forward model and the relevant noise properties. In principle, this gives access to the full information content of the data. In practice, however, the field-level approach is often viewed as tied either to specific perturbative descriptions, explicit reconstruction algorithms, or high-dimensional numerical implementations whose relation to more standard correlator-based analyses is not always transparent.

In this paper we develop a field-level posterior (FLP) for large-scale structure that makes this connection explicit, by directly expanding the FLP in terms of expectation values of correlators (see \cite{Cabass2024,Schmidt:2025iwa} and \cite{baumann/green,chen/green/lee} for related work). Our starting point is a general forward-model description in which the observed field is obtained from the initial conditions and a noise realization. Marginalizing over these latent variables yields a posterior directly for the evolved field. The construction is non-perturbative with respect to the forward model itself and therefore applies equally well when the correlators entering the description are obtained from perturbation theory, effective models, simulations, or more general data-driven approaches.

The main conceptual step is to expand the FLP around its Gaussian limit. Here,
``Gaussian'' refers to the statistics of the field, and does not necessarily imply linear-order perturbation theory as usually assumed in cosmology. That is, 
  the Gaussian part in principle already contains the fully nonlinear two-point information, while departures from Gaussianity are organized in terms of connected higher-order correlators. An example where such a reordering is beneficial is thermal field theory, where the thermal correction to the mass can be absorbed at all orders in a redefined free theory (see the discussion in Sect.~\ref{sec:VIexp}).
  
  Our expansion allows us to derive a general expression for the field-level Fisher matrix and to rewrite it as a series of contributions associated with the connected correlators of the evolved field. 
This organization makes it possible to compare field-level and correlator-based inference on the same footing. We show explicitly which terms in the field-level Fisher information are reproduced by likelihoods based on the power spectrum or the bispectrum, and how including cross-covariances among summary statistics
is important for consistency, i.e. to recover the correct match of information with the FLP.
In this sense, the FLP provides a unified language for quantifying the information loss induced by compression and for identifying which ingredients are needed for a given set of summaries to reproduce the field-level result.

A further motivation for this work is that the FLP also clarifies the status of ``reconstruction,'' or inference of the initial matter density field given the data. As an application, we discuss the baryon acoustic oscillation (BAO) scale and show how the full field contains the information needed for its optimal reconstruction in the presence of noise. Within the same formalism, the reconstruction of the initial field appears naturally as a byproduct. It is simply the optimal estimate of the initial conditions given the data and the noise model, thereby providing a direct bridge between field-level inference, correlator likelihoods, and reconstruction-based intuition.

Our formulation is related to earlier field-level constructions based on perturbative inversion of the forward model, or ``inverse perturbation theory,'' in which the final nonlinear field is expressed in terms of the linear one \cite{Cabass2024, Schmidt:2025iwa}. These works provide an important route to the FLP,
and in particular are directly related to practical FLP analyses relying on a field-level forward model \cite{Jasche:2012kq, Kitaura:2012tu, Wang:2013ep, Wang:2014hia, Modi:2018cfi, Schmidt:2020viy, Schmidt:2020tao, Shallue:2022mhf, Chen:2023uup, Andrews:2022nvv, Modi:2022pzm, Dai:2022dso, Kostic:2022vok, Qin:2023dew, Jindal:2023qew, Charnock:2019rbk, Doeser:2023yzv, Nguyen2024, Babic:2024wph}.
However, Refs.~\cite{Cabass2024, Schmidt:2025iwa} made simplifying assumptions that are not needed in the present framework. First, they are formulated in the noiseless limit, where the Gaussian likelihood connecting the data field to the nonlinear model collapses to a functional delta function. Second, implementing the perturbative inversion on a grid requires the initial and final fields to be represented on the same grid, and hence to carry the same number of degrees of freedom. In practice, this identifies the maximum wavenumber of the nonlinear field with that of the initial field. Here we instead keep these two scales distinct: the nonlinear field may be defined only up to an observationally relevant $\kmax$, while the ultraviolet cutoff entering perturbative loop integrals over the initial field can in general be much larger. The present formulation by design, does not rely on this identification of cutoffs.

A separate issue is the dependence of the discretized description on the grid. Achieving grid-size independence of observables requires a corresponding grid-size dependence of the field and model parameters, that is, a proper renormalization of the discretized theory \cite{Rubira:2023vzw,Rubira:2024tea,Peron:2025lgh}. In the present approach, the FLP is written as a series of convolutions of the observed field, $\delta_d$, with the connected correlators of the model. Its renormalization is therefore achieved once these correlators are properly renormalized, as is well understood, for example, within the EFTofLSS framework.

A further important aspect is the characterization of noise. Recent work has shown that non-Gaussian noise must be treated carefully in order to obtain unbiased field-level inference \cite{Akitsu:2025boy,Rubira:2025rqo} (see also \cite{Nguyen:2021} for a related earlier study). In our formulation, noise enters the FLP at the correlator level rather than through explicit realizations drawn from an assumed probability distribution, thereby making direct contact with existing EFTofLSS results for stochastic correlators (it is worth highlighting however that both correlators and field-level forward model involve the same expansion of stochastic contributions \cite{Rubira:2025rqo}).

The rest of the paper is organized as follows. In Sect.~\ref{sect:matter} we introduce the field-level posterior and its basic properties. In Sect.~\ref{sect:FLPFM} we derive the corresponding Fisher matrix and reorganize it into correlator contributions. In Sect.~\ref{sect:FMcorr} we compare these results with likelihoods for summary statistics. In Sect.~\ref{sect:FMres} we discuss the reconstruction of the BAO scale, while in Sect.~\ref{sect:reco} we show how the reconstruction of the initial field emerges within the same framework. We summarize our results and discuss possible extensions in Sect.~\ref{sect:concl}.

\section{The field-level posterior}
\label{sect:matter}
Our goal is to compute the field-level posterior,
\begin{align}
{\cal L}[\delta_d]=&\int  {\cal D}\delta_{\rm in}\, {\cal D} \epsilon\; \tilde{ \delta}_D\left[ \delta_d-\bar\delta[\delta_{\rm in},\epsilon]\right]\,{\cal P}[\delta_{\rm in},\,\epsilon]\,,
\label{eq:fllik}
\end{align}
where 
 ${\cal P}[\delta_{\rm in},\epsilon]$ denotes the joint probability distribution for the field initial conditions  and the noise (see \cite{Rubira:2025rqo} for a derivation of why a single noise field is sufficient to describe a single tracer in the EFTofLSS). We consider a finite cubic grid of size $L$ and $\Ng^3$ grid-node positions $\bx_i$. 
Our field variables are  {\em dimensionless}  fields in Fourier space,
\beq\delta_{\rm in}(\bm)\equiv \frac{1}{\Ng^3} \sum_i e^{-i \bk_\bm\cdot \bx_i}\delta_{\rm in}(\bx_i)\,,
\label{eq:FCadim}\eeq
with $\delta_{\rm in}(\bx_i)$  the configuration-space field, and analogous definition for the noise field in Fourier space, $\epsilon(\bm)$. The wavevector $\bk_\bm$ is expressed in terms of the fundamental grid mode $k_f\equiv 2\pi/L$ as
\beq
\bk_\bm\equiv  k_f\, \bm \,,
\eeq
where
\beq 
\bm=(m_x,m_y,m_z),\qquad{\rm and} \qquad  m_{x,y,z}=-\frac{\Ng}{2}+1,\cdots, \frac{\Ng}{2}\,,
\label{eq:rangen}
\eeq
excluding the zero mode $m_x=m_y=m_z=0$. Note that the grid size $\Ng$ provides a UV cutoff for all fields, $k_{\rm uv} \sim (\Ng/2) k_f$.
  We will assume that the model has been properly renormalized, so that all physical observables, more specifically, all  correlators of the field $\bar \delta[\delta_{\rm in},\epsilon]$, become independent of $\Ng$ on the scales of interest
(see below), or alternatively, that the model considered has a physical UV cutoff.\footnote{Note that the forward model might internally use a larger grid than $\Ng$, for example to avoid aliasing. Here, we set the initial conditions grid equal to the UV cutoff to avoid unnecessary notational complications.}
 
The corresponding integration measures are therefore defined as 
\beq
{\cal D}\delta_{\rm in}\equiv \Pi_{\bm}^{\Ng} \,d\delta_{\rm in}(\bm)\,,\qquad {\cal D}\epsilon\equiv \Pi_{\bm}^{\Ng} \,d\epsilon(\bm)\,,
\eeq
where the product is taken over the range \eqref{eq:rangen}.

$\bar\delta[\delta_{\rm in},\epsilon](\bm)$ represents the field generated by a forward model at redshift $z$, evolved from the initial conditions $\delta_{\rm in}(\bm)$ defined at an early time $z_{\rm in}$ and incorporating the stochastic noise field $\epsilon(\bm)$, while $\delta_d(\bm)$ is the data field, to be compared to the model. We restrict this comparison to modes with
  \beq
  |\bk_\bm| \leq \kmax\,.
\label{eq:kmax}
\eeq
While $\kmax \leq k_{\rm uv}$ is a physical requirement, in our derivation we need to make no further assumptions about a hierarchy between these two scales. 
Finally, $\tilde{\delta}_D\left[ \vp \right]$ indicates a functional delta function which enforces the $\kmax$ cut, defined as
\beq
\tilde{\delta}_D\left[ \vp \right]\equiv \Pi_{m}^{m_{\rm max}}\,\delta_D(\vp(\bn))\,,
\eeq
where the product of ordinary delta functions is taken over the wavenumbers satisfying Eq.~\eqref{eq:kmax}.

To disentangle the part of the FLP that originates from the model, we first recast \eqref{eq:fllik} in the form
\begin{align}
{\cal L}[\delta_d]=&\int {\cal D}\delta_{\rm M}\, {\cal D}\delta_{\rm in}\, {\cal D} \epsilon\; \tilde{\delta}_D\left[ \delta_d-\delta_{\rm M}\right]\,\tilde{\delta}_D\left[ \delta_{\rm M}-\bar\delta[\delta_{\rm in},\epsilon]\right]\,\,{\cal P}[\delta_{\rm in},\,\epsilon]\,,
\label{eq:fllikM} 
\end{align}
where the measure of the auxiliary field $\delta_{\rm M}(\bm)$ is
\beq
{\cal D}\delta_{\rm M}\equiv \Pi_{m}^{m_{\rm max}} \,d\delta_{\rm M}(\bm)\,,
\label{eq:measeps}
\eeq
with the product taken over the range \eqref{eq:kmax}, again excluding the zero mode (equivalently, this field could also be defined with the full measure  up to $\Ng$). At several points it is useful to distinguish clearly among the three fields appearing in the construction: $\delta_d$ denotes the observed data field, $\delta_{\rm M}$ the auxiliary nonlinear model field, and $\delta_{\rm in}$ the initial field.  We then express the first functional delta as
\beq
 \tilde{\delta}_D\left[ \delta_d-\delta_{\rm M}\right]= \int {\cal D} J_f\; e^{i  J_f\cdot\left(\delta_d-\delta_{\rm M}\right)} ,
\eeq
with
\beq
{\cal D} J_f\equiv  \Pi_{m}^{m_{\rm max}} \;\frac{d J(\bm)}{2 \pi}\,,\qquad 
J_f\cdot  \delta \equiv\sum_m^{m_{\rm max}} 
J_f(\bm) \delta(\bm)\,.
\eeq
Note that the support of $J_f$ is limited to those modes for which data and model are compared, i.e. Eq.~\eqref{eq:kmax}.
The FLP is then the functional Fourier transform of the generating functional,
\beq
{\cal L}[\delta_d]=\int {\cal D} J_f\; e^{i  J_f\cdot \delta_d} Z\left[ J_f\right]\,,
\label{eq:FTZ}
\eeq
where
\beq
Z\left[ J_f\right]\equiv \int {\cal D}\delta_{\rm M}\, e^{-i  J_f \cdot \delta_{\rm M}} {\cal P}_M[\delta_{\rm M}]\,,
\label{eq:ZJ}
\eeq
with
\beq
{\cal P}_M[\delta_{\rm M}]\equiv 
\int  {\cal D}\delta_{\rm in}\, {\cal D} \epsilon\;\,\tilde{\delta}_D\left[ \delta_{\rm M}-\bar\delta[\delta_{\rm in},\epsilon]\right]\,{\cal P}[\delta_{\rm in}, \epsilon]\,.
\eeq
It is straightforward to verify directly from Eq.~\eqref{eq:fllik} that \beq
 {\cal P}_M[\delta]={\cal L}[\delta]\,.
\label{eq:LeqP}
\eeq Nonetheless, we find it conceptually clearer to work first with the theoretical object $Z\left[J_f\right]$, evaluate it for a chosen forward model, and only at the end relate it to the data likelihood through Eq.~\eqref{eq:FTZ}.

\subsection{Expansion in connected functions}
We define the generating functional,
\beq
W[J_f]\equiv \ln Z[J_f]\,,
\eeq
from which we obtain the connected correlators of the evolved model field $\delta_{\rm M}$, including noise contributions,
\beq
i^j \left.\frac{\delta^j W[J_f] }{\delta J_f(\bm_1)\cdots \delta J_f(\bm_j)}\right|_{J_f=0}=\langle \delta_{\rm M}(\bm_1)\cdots \delta_{\rm M}(\bm_j) \rangle_c\equiv W_j(\bm_1,\cdots,\bm_j)\,.
\eeq
In this definition, we absorb into the dimensionless correlator $W_j$ the momentum-conserving factor $L^3 \delta_{\bm_{1,\cdots,j}}$, which equals $L^3$ for $\bm_{1,\cdots,j}\equiv \sum_{i=1}^j\bm_i=0$ and vanishes otherwise, as a consequence of translational invariance. This implies $W_1(\bm)=\langle \delta_{\rm M}(\bm) \rangle=0$. For $j=2$,  we use the notation
 \beq
 W_2(\bm)\equiv W_2(\bm,-\bm)\,.
 \eeq
The relation between our dimensionless correlators and the dimensional ones is given in Appendix \ref{app:conventions} (see Eqs.~\eqref{eq:WB} and \eqref{eq:appCnWn}).

The normalization of ${\cal P}_M[\delta_{\rm M}]$ ensures $Z[0]=1$ and therefore $W[0]=0$.  We can then express $Z[J_f]$ as
\beq
Z[J_f]=e^{W[J_f]}=e^{-\frac{1}{2}  J_f\cdot W_2\cdot  J_f} \; e^{V_I[J_f]}
\label{eq:Zsplit}
\eeq
where we isolate the term quadratic in the source, 
\beq
 J_f \cdot W_2 \cdot J_f\equiv  \sum_m^{m_{\rm max}} J_f(\bm )W_2(\bm) J_f(-\bm)\,,
\eeq
and expand the rest as 
\begin{align}
V_I[J_f]&=\sum_{j=3}^\infty \frac{(-i)^j}{j!}\sum_{m_1,\cdots,m_j}^{m_{\rm max}} \, W_j(\bm_1,\cdots \bm_j) J_f(\bm_1)\cdots  J_f(\bm_j)\,,\nonumber\\
&\equiv \sum_{j=3}^\infty \frac{(-i)^j}{j!}W_j \cdot J_f^j\,,
\label{eq:VI}
\end{align}
where the second line defines a short-hand notation we will use in the following, where contraction over $j$ implies sum over $\{\bm_1,\cdots \bm_j\}$.
Inserting Eq.~\eqref{eq:Zsplit} into Eq.~\eqref{eq:FTZ} and taking the interaction term outside the functional integral, we obtain
\begin{align}
{\cal L}[\delta_d]&=\int{\cal D}J_f \,e^{i J_f \cdot \delta_d}\,e^{-\frac{1}{2} J_f \cdot W_2 \cdot J_f} \; e^{V_I[J_f]}\nonumber\\
&= e^{V_I\left[-i\frac{\delta}{\delta\delta_d}\right]} {\cal L}_g[\delta_d]\,,
\label{eq:LL0}
\end{align}
where the Gaussian likelihood is 
\begin{align}
{\cal L}_{\rm g}[\delta_d]
&\equiv \int{\cal D}J_f \,e^{i J_f \cdot \delta_d}\,e^{-\frac{1}{2} J_f \cdot W_2 \cdot J_f}\nonumber\\ &=  \det\left[2 \pi W_2\right]^{-1/2}\,e^{-\frac{1}{2}\delta_d \cdot W_2^{-1}\cdot \delta_d}\,,
\nonumber\\
&=\exp\left\{-\frac{1}{2}\sum_{m}^{m_{\rm max}} \left[\frac{\left|\delta_d(\bm)\right|^2}{ W_2(\bm)} +\log \left(2 \pi W_2(\bm)\right)\right]\right\}\,,
\label{eq:L0gaussfull}
\end{align}
and the non-Gaussian part is encoded in the operator
\begin{align}
    V_I\left[-i \frac{\delta}{\delta\delta_d}\right]&=
 \sum_{j=3}^{\infty} \frac{\left(-1\right)^j}{j!}\sum_{m_1,\cdots,m_j}  W_j(\bm_1,\cdots,\bm_j) \frac{\delta^j}{\delta\delta_d(\bm_1)\cdots \delta\delta_d(\bm_j)}\,\,,\nonumber
 \\
 &\equiv \sum_{j=3}^{\infty} \frac{\left(-1\right)^j}{j!}  W_j \cdot \frac{\delta^j}{\delta \delta_d^j}\,.
\end{align}
Eqs.~\eqref{eq:LL0} and \eqref{eq:L0gaussfull} 
display how the information about the model, encoded in its connected correlators $W_j$, is propagated to the posterior probability of the field after marginalizing over both the initial conditions and the noise realizations. 
Note that $W_2$ represents the fully nonlinear power spectrum of the forward model, including the contribution from noise, and similarly $W_3$, $W_4$, etc.\ correspond to the fully nonlinear bispectrum, trispectrum, and higher-order spectra of the model. Up to this stage, no specific assumption about the forward model has been imposed, apart from translational invariance of the field $\delta_d$.
Although Eq.~\eqref{eq:LL0} is exact and non-perturbative, it naturally motivates an expansion in powers of the non-Gaussian contribution $V_I$, and, within that, an expansion in connected correlators starting from $j=3$. In what follows, we derive this double expansion.

\subsection{Expansion in \texorpdfstring{$V_I$}{VI}}
\label{sec:VIexp}

We expand Eq.~\eqref{eq:LL0}
as
\begin{align}
    {\cal L}= \sum_{k=0}^\infty
\frac{(V_I)^k}{k!} {\cal L}_g ={\cal L}_g  \sum_{k=0}^\infty
{\cal L}_g ^{-1}\frac{(V_I)^k}{k!} {\cal L}_g\,.
\label{eq:exp1}
\end{align}
The $k$th term in the expansion is
\begin{align}
{\cal L}_g ^{-1}\frac{(V_I)^k}{k!} {\cal L}_g&=\frac{1}{k!}\sum_{j_1,\cdots,j_k=3}^\infty \frac{(-1)^{j_1+\cdots +j_k}}{j_1!\cdots j_k!} {\cal L}_g ^{-1}\left(W_{j_1} \cdot \frac{\delta^{j_1}}{\delta \delta_d^{j_1}}\right)\cdots \left(W_{j_k} \cdot \frac{\delta^{j_k}}{\delta \delta_d^{j_k}}\right){\cal L}_g \,,\nonumber\\
&=\frac{1}{k!}\sum_{j_1,\cdots,j_k=3}^\infty \frac{1}{j_1!\cdots j_k!} \left[W_{j_1}\cdots W_{j_k}\right] \cdot {\cal H}_{j_1+\cdots +j_k}\,,
\label{eq:exp2}
\end{align}
where we have introduced the notation
\begin{align}
\left[W_{j_1}\cdots W_{j_k}\right]& \cdot {\cal H}_{j_1+\cdots +j_k}\nonumber\\ 
&\equiv \sum_{n_1^1,\cdots,n^1_{j_1}}^{m_{\rm max}}W_{j_1}(\bn^1_1,\cdots,\bn^1_{j_1})\cdots \sum_{n_1^k,\cdots,n_{j_k}^k}^{m_{\rm max}}W_{j_k}(\bn^k_1,\cdots,\bn_{j_k}^k)\nonumber\\&\qquad\times{\cal H}_{j_1+\cdots +j_k}(\bn^1_1,\cdots,\bn^1_{j_1},\cdots,\bn^k_1,\cdots,\bn^k_{j_k})\,,
\end{align}
and the  ${\cal H}_j[\delta_d]$'s are the functional generalizations of the probabilists' Hermite polynomials,
\begin{align}
{\cal H}_j[\delta_d](\bn_1,\cdots,\bn_j)
&\equiv (-1)^j\,{\cal L}_g[\delta_d] ^{-1} \frac{\delta^j}{\delta\delta_d(\bn_1)\cdots \delta\delta_d(\bn_j)}{\cal L}_g[\delta_d]\,.
\end{align}
${\cal H}_j[\delta_d]$ is a $j$-th order polynomial in the field $\delta_d$, whose coefficients are completely determined by $W_2(\bm)$. 
The first polynomials are
\begin{align}
&{\cal H}_0[\delta_d]=1\,\nonumber\\
&{\cal H}_1[\delta_d](\bm)=\frac{\delta_d(-\bm)}{W_2(\bm)}\,,\nonumber\\
&{\cal H}_2[\delta_d](\bm_1,\bm_2)=\frac{\delta_d(-\bm_1)}{ W_2(\bm_1)}\frac{\delta_d(-\bm_2)}{ W_2(\bm_2)}-\frac{\delta_{\bm_1+\bm_2}}{ W_2(\bm_1)}\,,\nonumber\\
&{\cal H}_3[\delta_d](\bm_1,\bm_2,\bm_3)=\frac{\delta_d(-\bm_1)}{ W_2(\bm_1)}\frac{\delta_d(-\bm_2)}{ W_2(\bm_2)}\frac{\delta_d(-\bm_3)}{ W_2(\bm_3)}-\frac{\delta_{\bm_1+\bm_2}}{ W_2(\bm_1)}\frac{\delta_d(-\bm_3)}{ W_2(\bm_3)}\nonumber\\
&\qquad\qquad\qquad\qquad\qquad-\frac{\delta_{\bm_2+\bm_3}}{ W_2(\bm_2)}\frac{\delta_d(-\bm_1)}{ W_2(\bm_1)}-\frac{\delta_{\bm_3+\bm_1}}{ W_2(\bm_3)}\frac{\delta_d(-\bm_2)}{ W_2(\bm_2)}\,.
\end{align}
Some useful properties of the ${\cal H}_j[\delta_d]$'s are collected in Appendix \ref{app:Hj}. 

In conclusion, the FLP~\eqref{eq:LL0}, expanded as in Eqs.~\eqref{eq:exp1} and \eqref{eq:exp2}, takes the form

\begin{align}    
{\cal L}[\delta_d]=&{\cal L}_g[\delta_d]\left(\sum_{k=0}^\infty\frac{1}{k!}\sum_{j_1,\cdots,j_k=3}^\infty \frac{1}{j_1!\cdots j_k!} \left[W_{j_1}\cdots W_{j_k}\right] \cdot {\cal H}[\delta_d]_{j_1+\cdots +j_k}\right)\,,\label{eq:exp3}\\
=&{\cal L}_g[\delta_d]\left(1+\sum_{k=1}^\infty\frac{1}{k!}\sum_{N=3}^{\infty} \frac{1}{N!}  \left[\tilde W_N^k\right]\cdot \mathcal{H}_N[\delta_d]\right)\,,
\label{eq:exp3Wien}
\end{align}
where
\begin{align}
    &\tilde W_N^k(\bn_1,\cdots,\bn_N)\equiv   \sum_{j_1,\cdots,j_k=3}^\infty \frac{N!}{j_1!\cdots j_k!}\,\delta_{j_1+\cdots+j_k-N} \nonumber\\
&\qquad\qquad\qquad\qquad\qquad\qquad\times W_{j_1}(\bn_1,\cdots,\bn_{j_1})\cdots W_{j_k}(\bn_{N-j_k+1},\cdots,\bn_N)\,.
\end{align}
Equation~\eqref{eq:exp3Wien} expresses the likelihood as a series expansion in Hermite functionals of increasing order, which is formally analogous to a functional Edgeworth expansion\footnote{ For previous uses of the Edgeworth expansion to describe non gaussianities in the CMB and in large scale structure see, for instance, \cite{Babich:2005en, Sellentin:2017aii, Philcox:2021ukg}.}. Such expansions are known to fail to preserve the positive definiteness of the PDF at extreme field values. In our setting, however, this approximation to the FLP remains well motivated because our primary aim is to compare field-level and correlator-based analyses in a regime where both are trustworthy, namely the  regime where the higher cumulants of the FLP are small. 

The expansion introduced here differs from the purely perturbative one because, at zeroth order, the likelihood is Gaussian but $W_2$ is identified with the fully nonlinear power spectrum, rather than with its linear counterpart. Within our framework, this choice emerges quite naturally and effectively corresponds to absorbing the complete non-Gaussian correction to the quadratic piece of the log-likelihood—i.e., the nonlinear power spectrum—into ${\cal L}_g[\delta_d]$. To begin with, as demonstrated in Appendix~\ref{app:corrLik}, employing the fully nonlinear $W_2$ in ${\cal L}_g[\delta_d]$ yields a Gaussian likelihood for the power spectrum that is centered on the nonlinear (instead of the linear) power spectrum, with a `Gaussian' covariance of $2 W_2^2$, where $W_2$ is the nonlinear kernel. Secondly, this procedure is entirely standard in Quantum Field Theory (QFT) whenever the physical degrees of freedom of interest differ from those of the non-interacting theory. As an illustration, take a scalar QFT in a thermal environment. A scalar field with bare mass $\mu$ and self-coupling $\lambda$ at temperature $T$ acquires a thermal mass correction, $\mu^2 \to \mu^2 + \lambda T^2/24$, due to forward scattering on the particles in the heat bath. It is then advantageous to rewrite the potential as
\begin{align}
V(\phi)=&\frac{\mu^2}{2}\phi^2+\frac{\lambda}{4}\phi^4\nonumber\\
=&\left(\frac{\mu^2}{2}\phi^2 + \frac{\lambda}{24}T^2\right)\phi^2  \nonumber\\
&\; - \frac{\lambda}{24}T^2 \phi^2+\frac{\lambda}{4}\phi^4\,,
\end{align}
where the first term in the last expression defines the new free (Gaussian) theory, and the remaining terms now play the role of the interaction, to be handled perturbatively.  
Although, to all orders, this reorganization is equivalent to the standard perturbative expansion in powers of $\lambda$, it is significantly more efficient because it better captures the actual physical degrees of freedom.

Coming back to our setting, a key question will be precisely to compare the expansion around the Gaussian FLP with a nonlinear power spectrum to the corresponding SPT expansion.

Our goal is to confront the FLP with likelihoods constructed from various summary statistics (see Sect.~\ref{sect:FMcorr}), all developed within the same theoretical framework and truncated at the same order in the correlators $W_j$, in order to assess how efficiently each method can extract information from the data. The Fisher matrix formalism offers a natural and self-consistent framework for carrying out this comparison.

\section{Field-level Fisher matrix}
\label{sect:FLPFM}

We now derive the Fisher matrix associated with the FLP and organize it into contributions labeled by the connected correlators entering the expansion of Sect.~\ref{sect:matter}.

Consider the first and second derivatives of the likelihood with respect to the model parameters $\alpha_a$ ($a=1, \cdots, N_{\rm par}$),
\begin{align}
\frac{\partial }{\partial \alpha_a}\left(-2 \log {\cal L}[\delta_d]\right)&=-\frac{2}{{\cal L} [\delta_d]} \frac{\partial }{\partial \alpha_a} {\cal L}[\delta_d]\,, \label{eq:d1L}\\
\frac{1}{2}\frac{\partial^2 }{\partial \alpha_a \,\partial\alpha_b}\left(-2 \log {\cal L}[\delta_d]\right)&= \frac{1}{{\cal L} [\delta_d]^2} \frac{\partial }{\partial \alpha_a} {\cal L}[\delta_d]\frac{\partial }{\partial \alpha_b} {\cal L}[\delta_d] - \frac{1}{{\cal L} [\delta_d]} \frac{\partial^2 }{\partial \alpha_a\,\partial \alpha_b} {\cal L}[\delta_d]\,. \label{eq:d2L}
\end{align}
We take the average of the above quantities assuming that the data field $\delta_d$ is distributed according to the fiducial model, using
\beq
\langle {\cal F}[\delta_d] \rangle_{\rm fid}\equiv \int {\cal D}\delta_d\, {\cal P}_{\rm M,\,fid}[\delta_d]\,{\cal F}[\delta_d]=
\int {\cal D}\delta_d\, {\cal L}_{\rm fid}[\delta_d]\,{\cal F}[\delta_d] \,,
\eeq
where ${\cal F}[\delta_d]$ is a generic functional of $\delta_d$, ${\cal P}_{\rm M,\,fid}[\delta_d]$  is the probability distribution of the field of the fiducial model, and in the last equality we have used Eq.~\eqref{eq:LeqP}.

The average of \eqref{eq:d1L} gives
\beq
\left\langle\frac{\partial }{\partial \alpha_a}\left(-2 \log {\cal L}[\delta_d]\right)\right\rangle_{\rm fid}= -2 \int {\cal D}\delta_d\, \frac{{\cal L}_{\rm fid}[\delta_d]}{{\cal L}[\delta_d]}\, \frac{\partial }{\partial \alpha_a} {\cal L}[\delta_d].
\eeq
When the model of the likelihood coincides with the fiducial one, that is, $\alpha_a=\alpha_a^{\rm fid}$ for all the parameters, and therefore  ${\cal L}[\delta_d]={\cal L}_{\rm fid}[\delta_d]$, we get
\beq
-2 \int {\cal D}\delta_d\, \, \frac{\partial }{\partial \alpha_a} {\cal L}_{\rm fid}[\delta_d]=-2 \frac{\partial }{\partial \alpha_a} \int {\cal D}\delta_d\,  {\cal L}_{\rm fid}[\delta_d]=0\,,
\eeq
where the last equality comes from the normalization condition $\int {\cal D}\delta_d\,  {\cal L}_{\rm fid}[\delta_d]=1$. Similarly, from \eqref{eq:d2L} we get the Fisher matrix element,
\beq
F_{ab}=\left\langle \frac{1}{2}\frac{\partial^2 }{\partial \alpha_a \,\partial\alpha_b}\left(-2 \log {\cal L}_{\rm fid}[\delta_d]\right) \right\rangle_{\rm fid} =\int {\cal D}\delta_d\,  \frac{1}{{\cal L}_{\rm fid} [\delta_d]} \frac{\partial }{\partial \alpha_a} {\cal L}_{\rm fid}[\delta_d]\frac{\partial }{\partial \alpha_b} {\cal L}_{\rm fid}[\delta_d]\,,\label{eq:FLFish}
\eeq
where the contribution from the second term in \eqref{eq:d2L} vanishes again by normalization. Equation~\eqref{eq:FLFish} thus expresses the Fisher matrix as a functional integral of the FLP and its first derivatives, evaluated on the fiducial values of the model parameters.

Using the first line of Eq.~\eqref{eq:LL0}, we get
\begin{align}
\frac{\partial }{\partial \alpha_a} {\cal L}_{\rm fid}[\delta_d]=&\int {\cal D}J_f\, e^{i J_f\cdot \delta_d -\frac{1}{2}J_f\cdot W^{\rm fid}_2 \cdot J_f+V^{\rm fid}_I[J_f]}\sum_{j=2}^\infty\frac{(-i)^j}{j!} \sum_{\bm_1\,\cdots,\bm_j}^{m_{\rm max}} \delta_{\bm_{1,\cdots, j}}\nonumber\\
&\qquad \times  J_f(\bm_1)\cdots J_f(\bm_j) \frac{\partial }{\partial \alpha_a}W^{\rm fid}_j(\bm_1,\cdots,\bm_j)\,,\nonumber\\
=&\sum_{j=2}^\infty\frac{(-1)^j}{j!}  \sum_{m_1\,\cdots,m_j}^{m_{\rm max}}\frac{\delta^j {\cal L}_{\rm fid}[\delta_d]}{\delta \delta_d(\bm_1)\cdots \delta \delta_d(\bm_j) }\frac{\partial }{\partial \alpha_a}W^{\rm fid}_j(\bm_1,\cdots,\bm_j)\,\nonumber\\
=&\sum_{j=2}^\infty\frac{(-1)^j}{j!} \frac{\partial }{\partial \alpha_a} W^{\rm fid}_j\cdot \frac{\delta^j {\cal L}_{\rm fid}[\delta_d]}{\delta \delta_d^j}\,,
\end{align}
where in the last line we have introduced a more compact notation.

The Fisher matrix \eqref{eq:FLFish} can then be expressed as the sum of contributions from different correlators of the model,
\begin{align}
F_{ab}=\sum_{i,j=2}^\infty F_{ab}^{(i,j)}\,,\label{eq:FMtot}
\end{align}
where, noting that the contribution $\propto (\partial^2/\partial\alpha_a\partial\alpha_b) W_j^{\rm fid}$ vanishes under the ensemble average and
using compact notation, we have, 
\begin{align}
F_{ab}^{(i,j)}\equiv \frac{(-1)^{i+j} }{i!\,j!} \left(\frac{\partial }{\partial \alpha_a} W^{\rm fid}_i \right)\cdot\left( \int {\cal D}\delta_d\,\frac{1}{{\cal L}_{\rm fid}[\delta_d]} \frac{\delta^i {\cal L}_{\rm fid}[\delta_d]}{\delta \delta_d^i}\frac{\delta^j {\cal L}_{\rm fid}[\delta_d]}{\delta \delta_d^j} \right)\cdot\left(\frac{\partial }{\partial \alpha_b} W^{\rm fid}_j\right)\,.\label{eq:FMpar}
\end{align}
Notice that although the full field-level Fisher matrix, Eq.~\eqref{eq:FLFish}, is explicitly positive definite, this property does not generally hold for its individual components, Eq.~\eqref{eq:FMpar}. This fact provides a useful diagnostic for assessing the reliability of the correlator expansion of the full FLP. From now on, to avoid clutter, we denote the fiducial FLP simply by ${\cal L}[\delta_d]$ whenever no confusion can arise.

\subsection{Power spectrum contribution}
\begin{figure}[t]
    \centering
    \includegraphics[width=1.\linewidth]{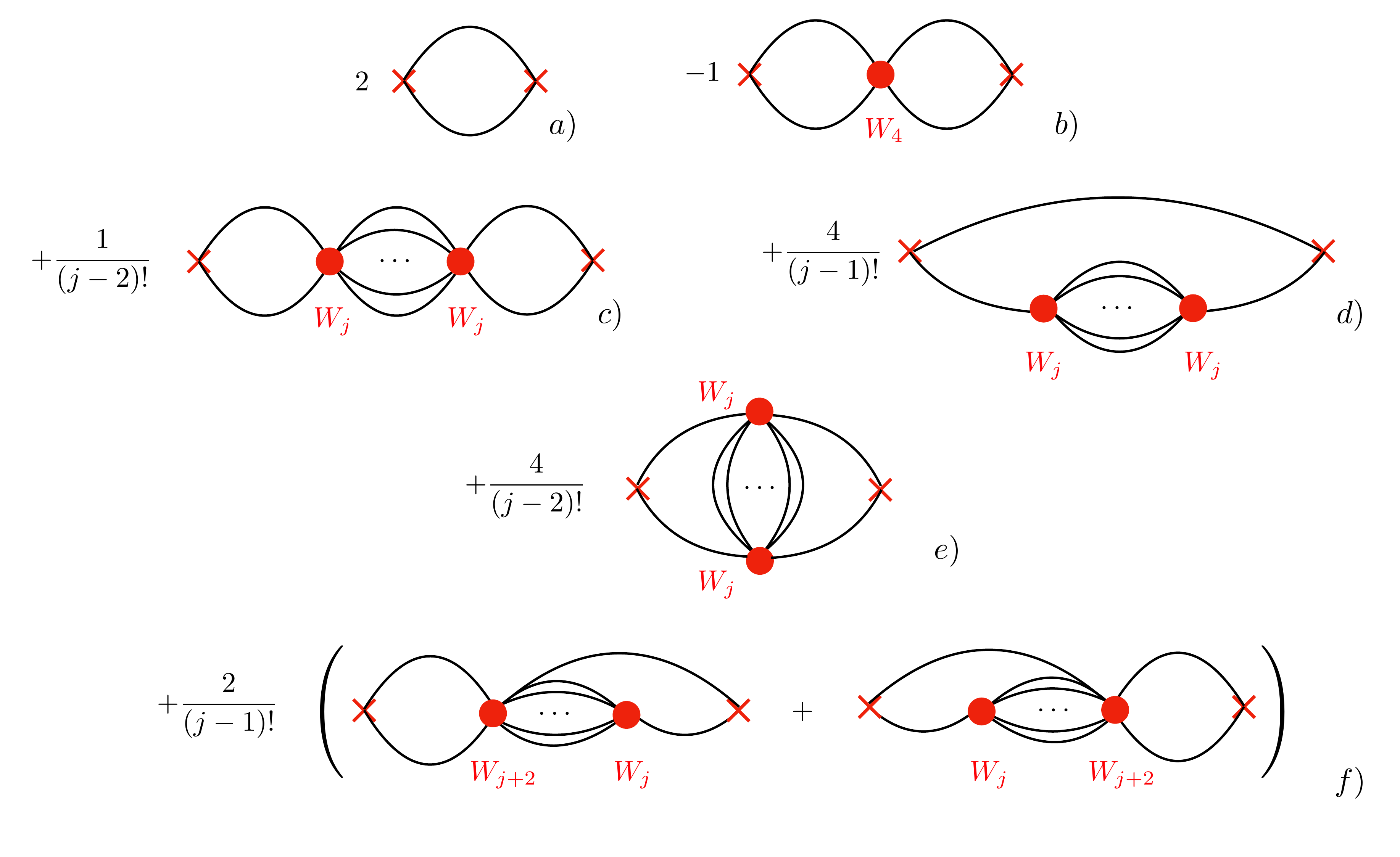}
    \caption{Diagrammatic illustration of the contributions to the Fisher matrix element $F_{ab}^{(2,2)}$ up to $O(V_I^2)$. The Gaussian contribution ($O(V_I^0)$) appears in diagram a). Diagram b) depicts the only non-Gaussian term at order $O(V_I)$, while diagrams c)–f) include all contributions at order $O(V_I^2)$. Note that all second-order contributions begin at $j = 3$, except for c), which is zero at $j = 3$ due to momentum conservation and therefore first contributes at $j = 4$. Note that the vertices $W_j$ here denote connected $j$-point functions, which can be taken to be fully nonlinear.}
    \label{fig:FMP2}
\end{figure}

The contribution to the Fisher matrix coming from the power spectrum $W_2$ is obtained by setting $i=j=2$ in Eq.~\eqref{eq:FMpar}.
Specifically, this is the Fisher matrix for parameters that only affect the power spectrum $W_2$, \emph{while keeping all higher-order $n$-point functions fixed.} We return to this point below. 
At second order in $V_I$ the result is given by (see Appendix \ref{app:FMcalc} for details),
\begin{align}
F_{ab}^{(2,2)}&=\frac{1}{4} \sum_{m_1,m_2}^{m_{\rm max}}\frac{\partial W_2(\bm_1) }{\partial \alpha_a}  \frac{\partial W_2(\bm_2) }{\partial \alpha_b}  \int {\cal D}\delta_d\,\frac{1}{{\cal L}[\delta_d]} \frac{\delta^2 {\cal L}[\delta_d]}{\delta \delta_d(\bm_1)\,\delta \delta_d(-\bm_1)}\frac{\delta^2 {\cal L}[\delta_d]}{\delta \delta_d(\bm_2)\,\delta \delta_d(-\bm_2)} \,,\nonumber\\
&=F_{ab,\text{g}}^{(2,2)}+F_{ab,\text{P{cov}}}^{(2,2)}+F_{ab,\text{FLP}}^{(2,2)}\,,\label{eq:FMps}
\end{align}
where
\begin{align}
F_{ab,\text{g}}^{(2,2)}= \frac{1}{2}\sum^{m_{\rm max}}_{m_1,m_2}\frac{\partial  W_2(\bm_1) }{\partial \alpha_a}  \frac{\partial  W_2(\bm_2) }{\partial \alpha_b}\frac{\delta_{\bm_1+\bm_2}}{W_2(\bm_1)^2}\,, 
\label{eq:FM22g}
\end{align}
is the Gaussian contribution, and
\begin{align}
F_{ab,\text{P{cov}}}^{(2,2)}=&  \frac{1}{4}\sum^{m_{\rm max}}_{m_1,m_2}\frac{\partial  W_2(\bm_1) }{\partial \alpha_a}  \frac{\partial  W_2(\bm_2) }{\partial \alpha_b}\Bigg[-\frac{W_4(\bm_1,-\bm_1,\bm_2,-\bm_2)}{ W_2(\bm_1)^2W_2(\bm_2)^2}\nonumber\\
&+\frac{1}{2} \sum_n^{m_{\rm max}}  \frac{ W_4(\bm_1,-\bm_1,\bn,-\bn)W_4(-\bn,\bn,\bm_2,-\bm_2)}{  W_2(\bm_1)^2 W_2(\bm_2)^2  W_2(\bn)^2}\Bigg]\,,
\label{eq:FM22NG}
\end{align}
is the contribution that is also present in the power-spectrum Fisher matrix; see Sect.~\ref{subsect:PSlik}. The remaining term,
\begin{align}
F_{ab,\text{FLP}}^{(2,2)}=& \frac{1}{4}\sum^{m_{\rm max}}_{m_1,m_2}\frac{\partial  W_2(\bm_1) }{\partial \alpha_a}  \frac{\partial  W_2(\bm_2) }{\partial \alpha_b}\frac{1}{W_2(\bm_1)W_2(\bm_2)} \Bigg\{\nonumber\\
&\sum_{j=5}^\infty\frac{1}{(j-2)!}\sum_{n_1\cdots n_{j-2}}^{m_{\rm max}} \frac{ W_j(\bn_1,\cdots,\bn_{j-2},\bm_1,-\bm_1)W_j(-\bn_1,\cdots,-\bn_{j-2},\bm_2,-\bm_2)}{  W_2(\bm_1) W_2(\bm_2)  W_2(\bn_1)\cdots W_2(\bn_{j-2})}\nonumber\\
&+\sum_{j=3}^\infty\Bigg[\frac{4\, \delta_{\bm_1+\bm_2}}{(j-1)!}\sum_{n_1\cdots n_{j-1}}^{m_{\rm max}}  \frac{W_j(\bn_1,\cdots,\bn_{j-1},-\bm_1)W_j(-\bn_1,\cdots,-\bn_{j-1},\bm_1)}{ W_2(\bm_1)   W_2(\bn_1)\cdots W_2(\bn_{j-1})}\nonumber\\
&+\frac{4}{(j-2)!} \sum_{n_1\cdots n_{j-2}}^{m_{\rm max}} \frac{W_j(\bn_1,\cdots,\bn_{j-2},-\bm_1,-\bm_2)W_j(-\bn_1,\cdots,-\bn_{j-2},\bm_1,\bm_2)}{ W_2(\bm_1)  W_2(\bm_2)  W_2(\bn_1)\cdots W_2(\bn_{j-2})}\nonumber\\
&+\frac{2}{(j-1)!}  \sum_{n_1\cdots n_{j-1}}^{m_{\rm max}}\Bigg( \frac{W_j(\bn_1,\cdots,\bn_{j-1},-\bm_1)W_{j+2}(-\bn_1,\cdots,-\bn_{j-1},\bm_1,\bm_2,-\bm_2)}{ W_2(\bm_1)W_2(\bm_2)   W_2(\bn_1)\cdots W_2(\bn_{j-1})}\nonumber\\
&\qquad\qquad \qquad\qquad \qquad\qquad +(\bm_1\leftrightarrow \bm_2)\Bigg)\Bigg]\Bigg\}\,,
\label{eq:FMPSFLP}
\end{align}
is a genuine new contribution present in the field-level Fisher matrix but not in the one for the power spectrum.
The various contributions can be depicted diagrammatically as in Fig.~\ref{fig:FMP2}. The term on the second line of Eq.~\eqref{eq:FMPSFLP} (diagram c)) begins at $j=5$ because the $j=3$ contribution vanishes (see below), while the $j=4$ term has already been included in the second line of Eq.~\eqref{eq:FM22NG}, since it corresponds to a contribution that also appears in the power spectrum Fisher matrix.

It is worth digressing briefly to discuss the $j=3$ contribution of diagram c). In the cubic volume with periodic boundary conditions considered here,
this contribution involves $W_3(0,\bm_1,-\bm_1)W_3(0,\bm_2,-\bm_2)$ via momentum conservation. Each factor vanishes because $\bar\delta({\bf 0})\equiv 0$ by assumption. If one considers a finite, non-periodic volume, the convolution with the corresponding window function in Fourier space would change this contribution to $W_3({\bf p},\bm_1,-\bm_1)W_3(-{\bf p},\bm_2,-\bm_2)$, where the momentum ${\bf p}$ is within the support of the window function (which we assume here is restricted to very large scales). One then recognizes this as the \emph{super-sample covariance} contribution \cite{takada/hu:2013}, which will correspondingly appear in the covariance of the power spectrum of the windowed density field, i.e. become part of Eq.~\eqref{eq:FM22NG}. Ref. \cite{barreira/etal:2018} presents a diagrammatic representation which clearly shows the resemblance to diagram c) in Fig.~\ref{fig:FMP2}, where the momentum on the line connecting the $W_3$ vertices is the large-scale ${\bf p}$.

In Sect.~\ref{subsect:PSlik} we compare this result with the Fisher matrix derived for the power spectrum under the assumption of a Gaussian likelihood. In particular, we will elucidate the origin of the contributions of Eq.~\eqref{eq:FMPSFLP}.
 As we will explain, the $O(W_3^2)$ terms capture the cross-correlation between the power spectrum and the bispectrum, and they appear in the power spectrum Fisher matrix only when it is derived from the joint power spectrum--bispectrum likelihood (which incorporates their cross-covariance), rather than from the likelihood for the power spectrum alone.  
The same reasoning applies to all the  terms in Eq.~\eqref{eq:FMPSFLP}, as well as to the corresponding new Fisher-matrix contributions arising from higher-order correlators.

This interpretation also helps identify the regime of validity of the expansion. Comparing the magnitude of the $O(W_4)$ correction in Eq.~\eqref{eq:FM22NG} to that of the Gaussian contribution, we find that the Fisher matrix is no longer positive definite when $W_4$ becomes large. More specifically, the relative size of the non-Gaussian correction scales as
\beq
 \sum^{m_{\rm max}}_{m_2} \frac{1}{2}\frac{W_4(\bm_1,-\bm_1,\bm_2,-\bm_2)}{ W_2(\bm_2)^2}\sim \int^{\kmax} \frac{d^3 p}{(2 \pi)^3} \frac{T(\bp,-\bp,\bk,-\bk)}{2 P(p)^2}\sim  \frac{\kmax^3}{8 \pi^2}\frac{T}{2 P^2},
 \eeq
where $T$ is the trispectrum and  $\kmax=2 \pi m_{\rm max}/L$.

\subsection{Bispectrum contribution}

We now consider the $i=j=3$ contribution to Eq.~\eqref{eq:FMpar}.
This corresponds to letting parameters that control $W_2$ and $W_3$ jointly vary, while keeping all higher order $(n\geq 4)$-point functions fixed.
Setting $i=j=3$ in Eq.~\eqref{eq:FMcov2} and inserting the result into
the former, we obtain the Fisher-matrix element
\beq
F_{ab}^{(3,3)}=F_{ab,\text{g}}^{(3,3)}+F_{ab,\text{B{cov}}}^{(3,3)}+F_{ab,\text{FLP}}^{(3,3)}\,.
\eeq 
\begin{figure}[t]
    \centering    \includegraphics[width=0.8\linewidth]{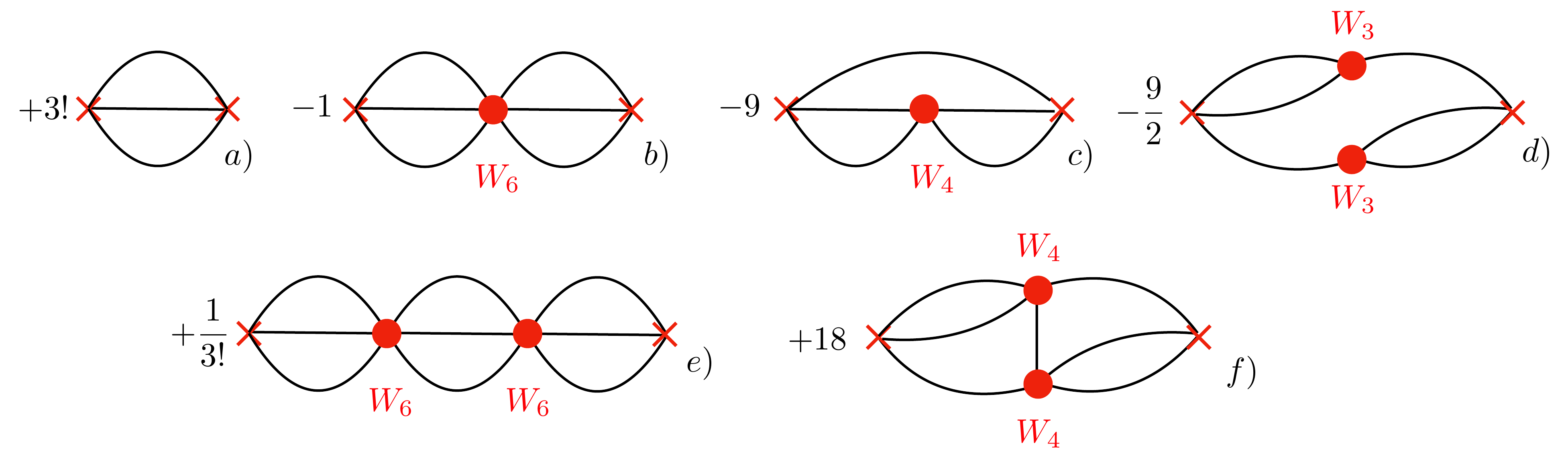}
\includegraphics[width=1.\linewidth]{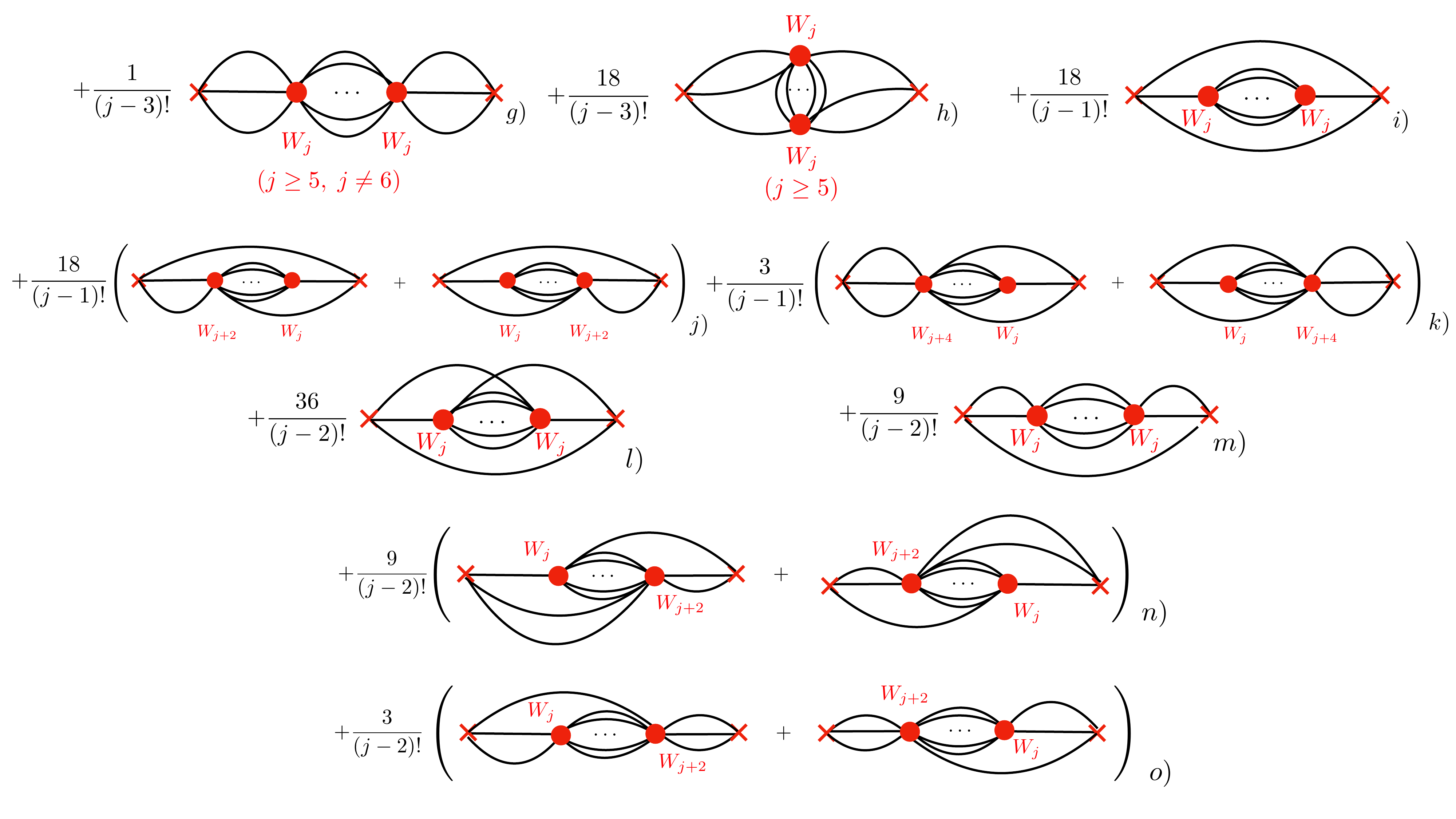}
    \caption{Diagrammatic illustration of the contributions to the Fisher matrix element $F_{ab}^{(3,3)}$ up to $O(V_I^2)$.}
    \label{fig:FMB2}
\end{figure}
The Gaussian contribution is 
\begin{align}
F_{ab, \text{g}}^{(3,3)}= \frac{1}{6}\sum_{m_1,m_2,m_3}^{m_{\rm max}} \frac{\left(\frac{\partial}{\partial\alpha_a}W_3(\bm_1,\bm_2,\bm_3)\right)\left(\frac{\partial}{\partial\alpha_b}W_3(\bm_1,\bm_2,\bm_3)\right)}{W_2(\bm_1)W_2(\bm_2)W_2(\bm_3)}\,,\label{eq:F33}
\end{align}
while the other two contributions are given in Eq.~\eqref{eq:bispV2} and represented diagrammatically in Fig.~\ref{fig:FMB2}.

Assuming a Gaussian likelihood for the bispectrum, the corresponding Fisher-matrix element is
\beq
F_{B,ab}^{(3,3)}=\sum_{t} \frac{\left(\frac{\partial}{\partial \alpha_a}B^{t}\right)\left(\frac{\partial}{\partial \alpha_b}B^{t}\right)}{{\rm Cov}_{B}^{t}}\,.\label{eq:Fc33}
\eeq
Here the sum runs over inequivalent triangle bins $t$, and ${\rm Cov}_{B}^{t}$ denotes the corresponding (full) bispectrum covariance. That is, $F_{B,ab}$ contains a part of $F_{ab,\rm Bcov}$ in addition to $F_{ab,g}$, but not the part that is due to the cross-covariance of power spectrum and bispectrum. As for the power spectrum, the relation between Eq.~\eqref{eq:Fc33} and the field-level result Eq.~\eqref{eq:F33} becomes more transparent in Sect.~\ref{sect:FMcorr} and in Appendix \ref{sect:FMbisp}, where the bispectrum likelihood is derived in the same notation.

\begin{figure}[t]
    \centering
    \includegraphics[width=1.\linewidth]{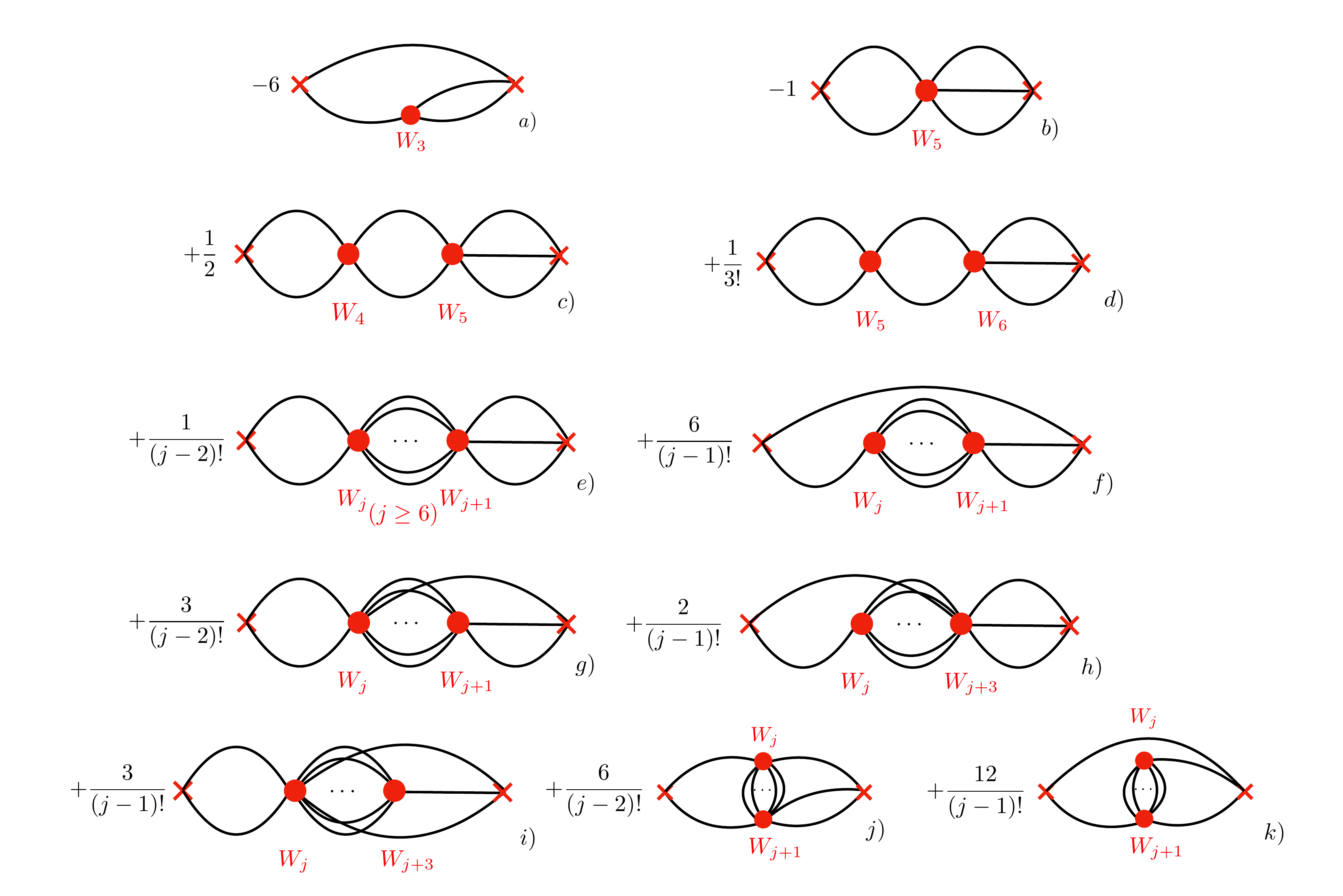}
    \caption{Diagrammatic illustration of the contributions to the Fisher matrix element $F_{ab}^{(2,3)}$ up to $O(V_I^2)$.}
    \label{fig:FMPB}
\end{figure}

The mixed contributions to $F_{ab}^{(2,3)}$ are shown in Fig.~\ref{fig:FMPB} and written explicitly in Eq.~\eqref{eq:PBVII}. 

\section{The likelihood and Fisher matrix for summary statistics}
\label{sect:FMcorr}

We now compare the field-level Fisher matrix with the Fisher matrices obtained after compressing the data to finite sets of summary statistics. The aim is to identify which terms in the FLP expansion are retained by a given compression and which are lost.

We denote by $\hat S[\delta]$ the estimator of a given summary statistic $S$ built from the field $\delta(\bn)$. In general, $\hat S[\delta]$ is a data vector. For correlators, for example, this vector is composed of the different power-spectrum bins, the bispectrum triangles, and so on, if higher-order correlators are included. The associated likelihood can be written compactly as a functional integral over the FLP,
\beq
{\cal L}_S[\hat S_d]=\int {\cal D}\delta_d \,\delta_D\!\left(\hat S_d-\hat S[\delta_d]\right)\, {\cal L}[\delta_d]\,.
\label{eq:likst}
\eeq
Note that, unlike in \eqref{eq:fllik}, where a functional delta function appears, the delta function in the expression above acts solely on the components of the data vector, rather than on the far more numerous modes or pixels on which the field is defined. Consequently, the mapping imposed by the delta function in Eq.~\eqref{eq:likst} is highly non-invertible and generally results in a significant compression of the data.

From this likelihood, we introduce the Fisher matrix associated with the summary statistics $S$ as
\begin{align}    
F^S_{ab}=&\left\langle \frac{1}{2}\frac{\partial^2 }{\partial \alpha_a \,\partial\alpha_b}\left(-2 \log {\cal L}_S[\hat S_d]\right) \right\rangle_{\rm fid} \,,\nonumber\\
=&\int {\cal D}\hat S_d\,  \frac{1}{{\cal L}^\text{fid}_S [\hat S_d]} \frac{\partial }{\partial \alpha_a}{\cal L}^\text{fid}_S [\hat S_d]\frac{\partial }{\partial \alpha_b} {\cal L}^\text{fid}_S [\hat S_d]\,,
\label{eq:FSab}
\end{align}
where, as in the previous discussion, we omit the superscript ``fid'' from now on whenever doing so does not lead to confusion.

Using Eq.~\eqref{eq:likst} we can relate this Fisher matrix to the one at field level. Indeed, the derivative with respect to the model parameter $\alpha_a$ can be expressed as
\begin{align}
    \frac{\partial {\cal L}_S[\hat S_d]}{\partial \alpha_a}&=\int {\cal D}\delta_d \,\delta_D\left(\hat S_d-\hat S[\delta_d]\right)\, \frac{\partial {{\cal L} [\delta_d]}}{\partial \alpha_a}\,,\nonumber\\
    &=\sum_{j=2}^\infty\frac{(-1)^j}{j!} \frac{\partial W_j}{\partial \alpha_a}\cdot \int {\cal D}\delta_d \,\delta_D\left(\hat S_d-\hat S[\delta_d]\right)\, \frac{\delta^j {{\cal L} [\delta_d]}}{\delta \delta_d^j}\,,\nonumber\\
    &=\sum_{j=2}^\infty\frac{(-1)^j}{j!} \frac{\partial W_j}{\partial \alpha_a}\cdot {{\cal L}_S^{(j)} [\hat S_d}]\,,
    \label{eq:LOder}
\end{align}
where
\beq
{\cal L}_S^{(j)}[\hat S_d](\bn_1,\cdots,\bn_j)\equiv \int {\cal D}\delta_d \,\delta_D\left(\hat S_d-\hat S[\delta_d]\right)\, \frac{\delta^j {{\cal L} [\delta_d]}}{\delta \delta_d(\bn_1)\cdots \delta \delta_d(\bn_j) }\,.
\eeq
We can then express the Fisher matrix \eqref{eq:FSab} as
\begin{align}
F^S_{ab}
&=\sum_{i,j=2}^\infty\frac{(-1)^{i+j}}{i!j!}\frac{\partial W_i}{\partial \alpha_a}\cdot\left( \int {\cal D}\hat S_d\,  {\cal L}_S^{(i)}[\hat S_d]\frac{1}{{\cal L}_S [\hat S_d]} {\cal L}_S^{(j)}[\hat S_d] \right)\cdot \frac{\partial W_j}{\partial \alpha_b}\,.
\label{eq:FabO}
\end{align}
The structural similarity to Eq.~\eqref{eq:FMpar} enables a clear comparison between the information carried by the full field and that encoded in a given summary statistic, represented by the data vector $\hat S_d$.

\subsection{Likelihood for the power spectrum} 
\label{subsect:PSlik}

To proceed, we consider the power spectrum. The estimator for the $i$-th spherical bin in momentum space is
\beq
\hat P^i_d=\hat P^i[\delta_d]=\frac{L^3}{ N_i}\sum_{\bn \in V_i}|\delta_d(\bn)|^2\,,
\label{eq:PSest}
\eeq
where the number of modes contained in the volume $V_i$ of the bin is given by\footnote{Note that only $1/2$ of the modes in the Fourier-space volume are actually independent, due to the reality constraint on $\delta_d$. In the literature, the Fourier-space volume is also often restricted to the half-space for this reason.}
\beq
N_i= \frac{L^3 k_i^2 \Delta k}{2 \pi^2}=4 \pi n_i^2 \Delta n\,,
\eeq
which is typically much larger than unity (it is $O(10^5)$ on the scales of interest for modern galaxy surveys).
Its expectation value can be computed using the FLP, as
\begin{align}    
\langle \hat P^i_d\rangle=\int {\cal D}\delta_d\,{\cal L}[\delta_d]\hat P^i[\delta_d]&= \frac{L^3}{ N_i}\sum_{\bn \in V_i} \int {\cal D}\delta_d\,{\cal L}[\delta_d]\,|\delta_d(\bn)|^2\,,\nonumber\\
&=\frac{L^3}{N_i}\sum_{\bn \in V_i} W_2(n)\equiv L^3 W_2^i\,,
\label{eq:psest1}
\end{align}
and is given by the average power spectrum in the bin considered. Notice that, in the correlator expansion of the FLP, Eq.~\eqref{eq:exp3Wien}, only the Gaussian $\mathcal{L}_g$ contributes, since 
\beq
\langle {\cal H}_N[\delta_d] |\delta_d|^2\rangle_g =0
\eeq for any $N\ge 3$, where $\langle\cdots\rangle_g$ indicates the average with respect to the Gaussian FLP.
The same result is obtained directly from the power-spectrum likelihood, using Eq.~\eqref{eq:likst},
\begin{align}
 \langle \hat P^i_d\rangle=\int d\hat P^i \,{\cal L}_{P^i}[\hat P^i]\,\hat P^i &=  \int  d\hat P^i \int\,{\cal D}\delta_d \,\delta_D\left(\hat P^i-\hat P^i[\delta_d]\right)\, {\cal L}[\delta_d] \hat P^i[\delta_d],\nonumber\\
 &=\int\,{\cal D}\,\delta_d {\cal L}[\delta_d] \hat P^i[\delta_d]\,,
\end{align}
which is identical to \eqref{eq:psest1}.

Considering $N_{\rm bin}$ bins, the power-spectrum estimator becomes a vector of $N_{\rm bin}$ elements $\{ \hat P_d^i\}$, and its likelihood becomes
\begin{align}
{\cal L}_{P}[\hat P_d]&=\Pi_{i=1}^{N_{\rm bin}}\,{\cal L}_{P^i}[\hat P_d^i]\,,\nonumber\\
&=\int {\cal D}\delta_d\,\Pi_{i=1}^{N_{\rm bin}}\delta_D(\hat P_d^i
-\hat P^i[\delta_d])\,{\cal L}[\delta_d]\,,\nonumber\\
&=\int {\cal D}\delta_d\,\int\left(\Pi_{i=1}^{N_{\rm bin}} \,\frac{dJ_i}{2\pi L^3}\right) e^{\frac{i}{L^3}\sum_{i=1}^{N_{\rm bin}} J_i\left(\hat P_d^i-\hat P^i[\delta_d]\right) }{\cal L}[\delta_d]\,,\nonumber\\
&=\int {\cal D}J\, e^{i J \cdot \hat P_d} \,Z_P[J]\,,
\label{eq:LOZ}
\end{align}
where we have introduced a vector of sources, $\{J_i\}$, and the compact notation
\beq
{\cal D}J\equiv \left(\Pi_{i=1}^{N_{\rm bin}} \,\frac{dJ_i}{2\pi L^3}\right)\,, \quad J \cdot \hat P_d \equiv \frac{1}{L^3}\sum_{i=1}^{N_{\rm bin}} J_i P_d^i\,.
\eeq
We also define the generating functional for the power spectrum,
\beq
Z_P[J]\equiv e^{W_P[J]}\equiv\int {\cal D}\delta_d\,\, e^{-i J\cdot \hat P[\delta_d]}\,{\cal L}[\delta_d]\,.
\label{eq:ZOJ}
\eeq
The first functional derivative gives the expectation value of the power spectrum,
\begin{align}
i\left.\frac{\delta W_P[J]}{\delta J_i}\right|_{J=0}&=L^{-3}\int{\cal D}\delta_d\,{\cal L}[\delta_d] \hat P^i[\delta_d]=W_2^i,    
\end{align}
and the second derivative gives the power spectrum covariance,
\begin{align}
    i^2 \left.\frac{\delta^2 W_P[J]}{\delta J_i\, \delta J_j}\right|_{J=0}&\equiv \Sigma^{ij}\,,\nonumber\\
    &=L^{-6}\int{\cal D}\delta_d\,{\cal L}[\delta_d] \hat P^i[\delta_d]\,\hat P^j[\delta_d] - W_2^i \,W_2^j\,,\nonumber\\
    &=\frac{1}{ N_i N_j}\sum_{\bn \in V_i,\, \bm\in V_j} \int {\cal D}\delta_d \, {\cal L}[\delta_d]|\delta_d(\bn)|^2|\delta_d(\bm)|^2 - W_2^i\,W_2^j\,.
\end{align}
To compute the first term in the last line we employ again the expansion \eqref{eq:exp3Wien}. The Gaussian approximation to the FLP gives
\beq
\frac{1}{ N_i N_j}\sum_{\bn \in V_i,\, \bm\in V_j} \int {\cal D}\delta_d \, {\cal L}_g[\delta_d]|\delta_d(\bn)|^2|\delta_d(\bm)|^2-\,W_2^i W_2^j= \frac{2 \delta_{ij}}{N_i^2}\sum_{\bn \in V_i} (W_2(\bn))^2=\frac{2 \delta_{ij}}{N_i} (W_2^i)^2\,.
\eeq
Among the non-Gaussian contributions to Eq.~\eqref{eq:exp3Wien}, only the term linear in $W_4$ contributes,
\begin{align}
&\frac{1}{ N_i N_j}\sum_{\bn \in V_i,\, \bm\in V_j}\frac{1}{4!}\langle [W_4]\cdot {\cal H}_4[\delta_d] |\delta_d(\bn)|^2|\delta_d(\bm)|^2 \rangle_g\,,\nonumber\\
&=\frac{1}{N_i N_j}\sum_{\bn \in V_i,\, \bm\in V_j} W_4(\bn,-\bn,\bm,-\bm)= \,W_4^{ij}\,,
\end{align}
where we have defined the bin-averaged trispectrum as
\beq
W_4^{ij}\equiv \frac{1}{N_i N_j} \sum_{\bn \in V_i,\, \bm\in V_j} W_4(\bn,-\bn,\bm,-\bm)\,.
\eeq
Therefore, the full power spectrum covariance is 
\beq
\Sigma^{ij}=\frac{2 }{N_i} (W_2^i)^2 \,\delta_{ij}+ \,W_4^{ij}\,. 
\label{eq:Sigma}
\eeq
We now expand the generating functional $W_P[J]$ as (noticing that $W_P[0]=0$ due to the normalization of ${\cal L}[\delta_d]$)
\begin{align}
    W_P[J]=&\sum_{i=1}^{N_{\rm bin}}\left.\frac{\delta W_P[J]}{\delta J_i}\right|_{J=0} J_i+\frac{1}{2} \sum_{i,j=1}^{N_{\rm bin}}\left.\frac{\delta^2 W_P[J]}{\delta J_i\delta J_j}\right|_{J=0} J_i J_j+\cdots\,,\nonumber\\
    =& -i \,J\cdot W_2 -\frac{1}{2}\,J\cdot \Sigma\cdot J+\cdots\,.
    \label{eq:WPexp}
\end{align}
Since we are interested in comparing the Fisher matrix from ${\cal L}_P[\hat P_d]$ with the one obtained from the FLP, we truncate the expansion at quadratic order in the sources $J_i$. See Appendix~\ref{app:corrLik} for a discussion of higher orders and for an assessment of the accuracy of the Gaussian approximation to ${\cal L}_P[\hat P_d]$ in the large $N_i$ limit.

\begin{figure}[t]
    \centering
    \includegraphics[width=1.\linewidth]{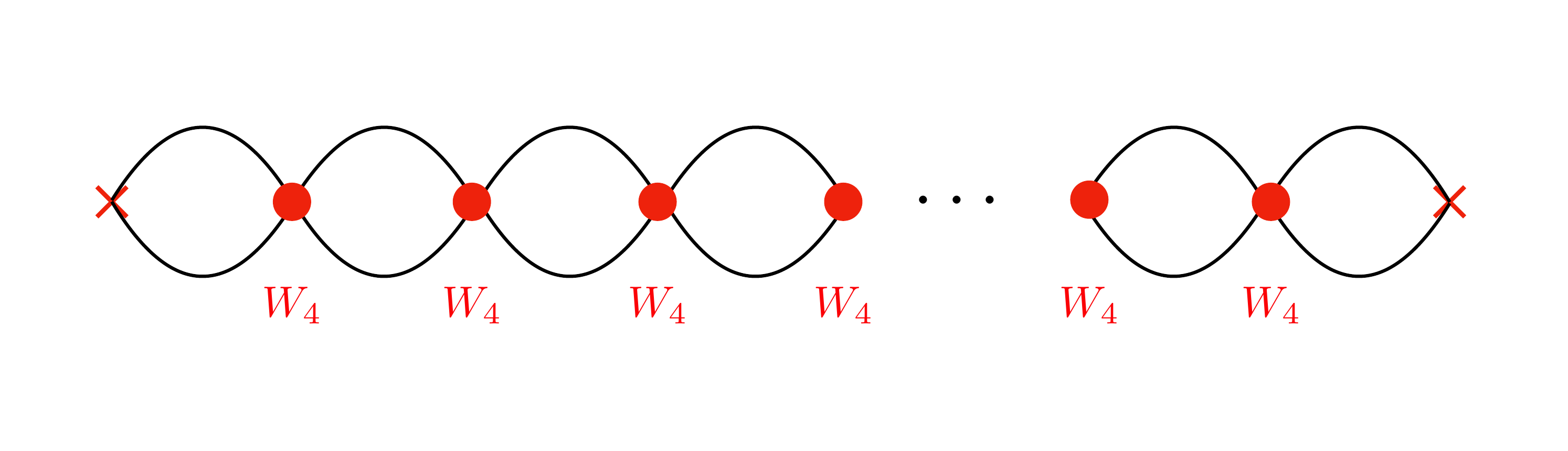}
    \caption{Higher-order contributions to the power spectrum Fisher matrix, obtained by the expansion in Eq.~\eqref{eq:PScovinv}.}
    \label{fig:FMPSres}
\end{figure}

By inserting Eq.~\eqref{eq:WPexp} into Eq.~\eqref{eq:LOZ} and performing the integration over the sources, we obtain the Gaussian approximation to the power spectrum likelihood,
\begin{align}
{\cal L}_P[\hat P_d]\simeq& \frac{1}{\sqrt{(2 \pi)^{N_{\rm bin}} \det(L^6\Sigma)}}\,\exp\!\left[-\frac{1}{2}\left(\hat P_d L^{-3}-W_2\right)\cdot \Sigma^{-1}\cdot \left(\hat P_d L^{-3}-W_2\right)\right]\,,
\label{eq:LPsgauss}
\end{align}
from which the Fisher matrix follows as
\beq
 F_{P;ab}=  \sum^{N_{\rm bin}}_{i,j=1}\frac{\partial W_2^i }{\partial \alpha_a}  \frac{\partial W_2^j }{\partial \alpha_b} \,\Sigma_{ij}^{-1}\,.
\label{eq:FPS}
\eeq
Inverting the covariance in Eq.~\eqref{eq:Sigma} and expanding in powers of $W_4$, we find
\begin{align}
  \Sigma_{ij}^{-1}=&  \frac{\delta_{ij}}{2 (W_2^i)^2} N_i-\frac{1}{4} N_iN_j \frac{W_4^{ij}}{ (W_2^i)^2 (W_2^j)^2}\nonumber\\
  &+\frac{1}{8}N_i N_j\sum_{l=1}^{N_{\rm bin}} N_l \frac{W_4^{il}W_4^{lj}}{ (W_2^i)^2(W_2^l)^2 (W_2^j)^2}+ O(W_4^3)\,.
  \label{eq:PScovinv}
\end{align}
The leading term, when substituted into Eq.~\eqref{eq:FPS}, reproduces the Gaussian part of the “2-2” block of the Fisher matrix obtained from the FLP, Eq.~\eqref{eq:FM22g},
\begin{align}
F_{ab,\text{g}}^{(2,2)}=& \frac{1}{2}\sum^{m_{\rm max}}_{m,n}\frac{\partial  W_2(\bm) }{\partial \alpha_a}  \frac{\partial  W_2(n) }{\partial \alpha_b}\frac{\delta_{\bm+\bn}}{W_2(\bm)^2}\,, \nonumber\\
=& \frac{1}{2}\sum^{N_{\rm bin}}_{i,j=1} \;\;\sum_{\bm\in V_i, \,\bn\in V_j}\frac{\partial  W_2(\bm) }{\partial \alpha_a}  \frac{\partial  W_2(n) }{\partial \alpha_b}\frac{\delta_{\bm+\bn}}{W_2(\bm)^2}\,,\nonumber\\
=&\frac{1}{2}\sum^{N_{\rm bin}}_{i,j=1}\delta_{ij} \;\;\sum_{\bm\in V_i} \frac{\partial  W_2(\bm) }{\partial \alpha_a}  \frac{\partial  W_2(\bm) }{\partial \alpha_b}\frac{1}{W_2(\bm)^2}\,,\nonumber\\
\simeq&\sum^{N_{\rm bin}}_{i,j=1}\delta_{ij}  \frac{\partial W_2^i}{\partial \alpha_a}\frac{\partial W_2^i}{\partial \alpha_b}\frac{N_i}{2(W_2^i)^2 }\,,
\end{align}
where we have replaced the power spectrum (and its derivatives) within each bin with its bin-averaged value,
$$ W_2(\bm)\simeq W_2^i\qquad \text{for} \;\;\bm\in V_i\,.$$
Similarly, the $O(W_4)$ and $O(W_4^2)$ contributions in Eq.~\eqref{eq:PScovinv} can be derived from Eq.~\eqref{eq:FM22NG} by approximating the mode-level power spectrum and $W_4$ with their bin-averaged counterparts. These three contributions are represented by the first two diagrams in Fig.~\ref{fig:FMP2}, and by the third diagram for $j=4$. Continuing the series in Eq.~\eqref{eq:PScovinv} generates chain-type diagrams of all orders in $W_4$, as depicted in Fig.~\ref{fig:FMPSres}.

In summary, the Fisher matrix for the power spectrum corresponds to the sum of Eqs.~\eqref{eq:FM22g}, \eqref{eq:FM22NG}, and the chain diagrams of Fig.~\ref{fig:FMPSres},
\beq
 F_{P;ab}\simeq F_{ab,\text{g}}^{(2,2)} + F_{ab,\text{Pcov}}^{(2,2)} + \text{``$W_4$ chain diagrams''}\,.
\eeq

In contrast, the remaining contributions to the “2-2” field-level Fisher matrix, indicated as $F_{ab,\text{FLP}}^{(2,2)}$ in Eq.~\eqref{eq:FMPSFLP} (and corresponding to diagram c) with $j>4$ and diagrams d)–f) in Fig.~\ref{fig:FMP2}), have no counterpart in the power-spectrum Fisher matrix. To capture them, the correlator data vector must be enlarged.

  It might be surprising at first that higher-order summary statistics are important, given that, by assumption, the parameter vector considered only modifies $W_2(\bm)$. This is due to the cross-covariance of $W_2$ with higher-order correlation functions, which are implicitly kept fixed when considering the Fisher information. As demonstrated in App.~\ref{app:fishermarg}, for a parameter $p_\alpha$ that affects only the power spectrum, the information obtained about it from the power spectrum and from higher-order statistics becomes equivalent once we marginalize over the parameters governing the latter. Importantly, this no longer holds if parameters modify both $W_2$ and higher-order statistics, in which case the latter will add genuine new information.

\subsection{Likelihood for P+B}
We now enlarge the data vector by adding the bispectrum estimator to the power-spectrum estimator \eqref{eq:PSest}:
\beq
\hat B^{t}_d=\hat B^{t}[\delta_d]=\frac{L^6}{ N_T^{t}}\sum_{\bm_1 \in V_i}\sum_{\bm_2 \in V_j}\sum_{\bm_3 \in V_k}\delta_{\bm_{123}}\,\delta_d(\bm_1) \delta_d(\bm_2) \delta_d(\bm_3)\,,
\eeq
where $t$ identifies the set of closed triangles with side lengths in the momentum bins $V_{i,j,k}$, and $N_T^{t}$ denotes the total number of triangles in the set $t$.

Following the same steps as for the power spectrum alone, we obtain the Gaussian likelihood for the joint ``P+B'' data vector,
\beq
V_{PB}^\alpha \equiv \{P_d^i L^{-3},\, B_d^t L^{-6}\}\,,
\eeq
where $\alpha = i,\, j,\, k, \cdots$ labels the power spectrum bins, and $\alpha = t,\, u,\, v, \cdots$ labels the binned triangles.
The covariance matrix now has the block form
\begin{align}
\Sigma_{P+B}^{\alpha\beta}=\left(
\begin{array}{cc}
\Sigma^{ij} & \Sigma_{PB}^{iu} \\
\Sigma_{BP}^{tj} & \Sigma_B^{tu}
\end{array}
\right)\,,
\end{align}
with
\begin{align}
\Sigma_B^{tu}=\delta_{tu}\Sigma_{g}^{t}+\Sigma_{W_3^2}^{tu}+\Sigma_{W_2 W_4}^{tu}+\Sigma_{W_6}^{tu}\,.
\end{align}
Here $\Sigma_g^t$ is the Gaussian bispectrum covariance, cubic in the power spectrum, while the remaining terms arise respectively from $W_3^2$, $W_2W_4$, and $W_6$.

The new ingredient relative to the power-spectrum-only case is the mixed covariance between the power spectrum and the bispectrum. At leading order it is generated by the disconnected contraction of five fields and scales as
\beq
\Sigma_{PB}^{i t}\propto W_2^i W_3^{t}\,,
\eeq
where $t$ labels the triangle configuration. Upon inverting the block covariance matrix, the power spectrum block of the inverse covariance is given by the Schur complement,
\begin{align}
\left(\Sigma_{P+B}^{-1}\right)_{PP}=&\,\Sigma^{-1}
+\Sigma^{-1}\Sigma_{PB}\left(\Sigma_B-\Sigma_{BP}\Sigma^{-1}\Sigma_{PB}\right)^{-1}\Sigma_{BP}\Sigma^{-1}\,.
\label{eq:PBschur}
\end{align}
Expanding Eq.~\eqref{eq:PBschur} to leading order in the mixed covariance gives
\begin{align}
\left(\Sigma_{P+B}^{-1}\right)_{PP}
\simeq \Sigma^{-1}+\Sigma^{-1}\Sigma_{PB}\Sigma_{B,g}^{-1}\Sigma_{BP}\Sigma^{-1}+\cdots\,.
\end{align}
Therefore the power-spectrum entry of the Fisher matrix receives the extra contribution
\begin{align}
    F_{P+B;ab}=F_{P;ab}+\frac{1}{4}\sum_{i,j=1}^{N_{\rm bin}} \frac{\partial W_2^i}{\partial\alpha_a}\frac{\partial W_2^j}{\partial\alpha_b}\frac{N_i N_j}{(W_2^i)^2 (W_2^j)^2}\sum_{t}\frac{\Sigma_{PB}^{i t}\Sigma_{BP}^{t j}}{ \Sigma_g^{t}}\,,
\end{align}
where $\Sigma_g^{t}$ is the Gaussian bispectrum covariance for triangle bin $t$. This reproduces the $j=3$ contribution in the third and fourth lines of Eq.~\eqref{eq:FMPSFLP}, namely the first non-trivial FLP correction to the power-spectrum block that cannot be captured by a power-spectrum-only likelihood.

Analogously, the BB block of $\Sigma_{P+B}^{-1}$ gives the bispectrum contribution to the Fisher matrix. Its leading term, $\Sigma_{B,g}^{-1}$, reproduces the Gaussian BB entries, namely Eq.~\eqref{eq:Fc33} and thus Eq.~\eqref{eq:F33}, while the non-Gaussian terms in $\Sigma_B^{tu}$, starting with $\Sigma_{W_3^2}^{tu}$, generate the corresponding non-Gaussian corrections. In this sense, Eq.~\eqref{eq:F33} can also be viewed as arising from the Gaussian part of a likelihood for the bispectrum, and more generally from the BB sector of a likelihood for an infinite hierarchy of correlators.

The same logic extends to higher correlators. Enlarging the data vector to include the trispectrum, the five-point function, and so on, progressively reproduces the remaining terms in the FLP expansion through the corresponding covariance blocks and cross-covariances. The key point is therefore that the Fisher information encoded in the FLP can be reorganized as the information contained in a Gaussian likelihood for an infinite hierarchy of correlators, provided their full covariance matrix---including all cross-correlations---is retained.

Our calculation thus shows that incorporating the cross-covariance between power spectrum and bispectrum is necessary to obtain a consistent posterior. In general, neglecting the cross-covariance could lead to either over- or underestimation of inferred parameter errors, depending on the detailed structure of the joint covariance as well as the parameter dependence of power spectrum and bispectrum.

\subsection{Other summary statistics}
\label{sect:FMss}
When dealing with summary statistics other than correlators, a natural question is how to interpret their information content in the language of correlators.
Eq.~\eqref{eq:exp3} offers a natural quantitative framework to address this. Specifically, for a chosen fiducial cosmology, the expectation value of a given summary statistic can be written as
\begin{align}
    \langle S[\delta]\rangle^{\rm fid}_{\cal L}=&\langle S[\delta]\rangle^{\rm fid}_{{\cal L}_g} + \sum_{k=1}^\infty\frac{L^{3k}}{k!}\sum_{N=3}^\infty\left[W_N^k\right]\cdot\langle {\cal H}_N[\delta]\,S[\delta]\rangle^{\rm fid}_{{\cal L}_g}\,\nonumber\\
    =&\langle S[\delta]\rangle^{\rm fid}_{{\cal L}_g}+\sum_{k=1}^\infty\frac{L^{3k}}{k!}\sum_{N=3}^\infty\left[W_N^k\right]\cdot{\cal C}_N\,,
\end{align}
where
\beq
{\cal C}_N(\bn_1,\cdots,\bn_N)\equiv \langle {\cal H}_N[\delta](\bn_1,\cdots,\bn_N)\,S[\delta]\rangle^{\rm fid}_{{\cal L}_g}\,,
\eeq
denotes the $N$-th order non-Gaussian weight associated with the summary statistic 
$S[\delta]$. 

Here the average is taken with respect to the Gaussian FLP. Thus these weights can be computed from Gaussian realizations of the field $\delta$, followed by an evaluation of the corresponding correlators.

\section{Recovering the BAO scale}
\label{sect:FMres}
In this section we illustrate how the FLP framework captures the recovery of the BAO scale once information carried by higher-order correlators is included.
Precisely, we show how the joint dependence of $W_2$ and $W_3$ on the BAO scale effectively captures the leading additional information that is obtained with BAO reconstruction techniques, where ``leading'' refers to the impact of modes much larger than the BAO scale.

When an additional physical scale is present, the standard perturbation theory (PT) power counting can break down: higher-order PT terms may become comparable in size to lower-order ones and must therefore be retained in the calculation. A classic and phenomenologically important case is the BAO scale, denoted by $r_d$, which shows up in the oscillatory, or ``wiggly,'' component of the nonlinear power spectrum,
\beq
W_2(k)=W_{2,\text{nw}}(k)+W_{2,\text{w}}(k),
\eeq
with
\beq
W_{2,\text{w}}(k)= A_{\text w}(k) \sin(k r_d)\,,
\eeq
where $W_{2,\text{nw}}(k)$ and $A_\text{w}(k)$ are smooth on the characteristic scale $r_d^{-1}$, in the sense that
\beq
\frac{d \log W_{2,\text{nw}}}{d \log k}\ll k r_d\,,\qquad \frac{d \log A_\text{w}}{d \log k}\ll k r_d\,.
\eeq
In this section we work in the continuum limit for the correlators, i.e. we assume $ q L /(2 \pi)\gg 1$ for all relevant modes $q$, and therefore  sums over discrete wavevectors are replaced by integrals.

The couplings between long-wavelength modes $q\ll k$ and modes with $k> 2\pi/r_d$ lead to a damping of the oscillatory part of the power spectrum,

\beq
A_\text{w}(k)=A^0_\text{w}(k) e^{-k^2\Sigma_\Lambda(r_d)}\,,
\eeq
where, assuming that the long modes are linear, the damping exponent is  
\beq
- k^2 \Sigma_\Lambda(r_d)=- k^2 \int^\Lambda
\frac{d^3 q}{(2\pi)^3}  \frac{P(q)}{q^2} \mu^2 \left(1-\cos(q\mu r_d)\right),
\eeq
with $P(q)$ the (dimensional) linear matter power spectrum, $\mu\equiv (\bk\cdot \bq )/(k q)$, and the integral restricted to $q\le \Lambda \ll k$. By splitting the integration domain, this integral can be approximated as
\beq
k^2 \Sigma_\Lambda(r_d)\simeq k^2 \sigma_{\Lambda,v}^2 + \frac{k^2 r_d^2}{10}\Delta^2_{r_d}\,,
\eeq
where the contribution from large-scale bulk flows is
\beq
\sigma_{\Lambda,v}^2\equiv \frac{1}{3}\int_{2\pi r_d^{-1}}^\Lambda \frac{d^3 q}{(2\pi)^3}\frac{P(q)}{q^2}\,,
\label{eq:sigmav}
\eeq
and the contribution from long-wavelength tidal fields is
\beq
\Delta^2_{r_d}\equiv \int^{2\pi r_d^{-1}}\frac{d^3 q}{(2\pi)^3} P(q)\,.
\eeq
We now define the parameter 
\beq
\beta\equiv \log\left(\frac{r_d}{r_d^\text{fid}}\right)\,,
\eeq
where $r_d^\text{fid}$ is the fiducial BAO drag scale. We then compute the Fisher matrix for $\beta$. The Gaussian power-spectrum contribution (diagram a in Fig.~\ref{fig:FMP2}) yields
\begin{align}
    F_{\beta,g}^{(2,2)}=&L^3\int \frac{d^3 k}{(2\pi)^3} \left(\frac{\partial W_2(k)}{\partial \beta}\right)^2\frac{1}{2 W_2(k)^2 }\,,\nonumber\\
    \simeq&L^3\int \frac{d^3 k}{(2\pi)^3} \left(\frac{\partial W_{2,\text{w}}(k)}{\partial \beta}\right)^2\frac{1}{2 W_2(k)^2 }\,,\nonumber\\
    \simeq&L^3 \int \frac{d^3 k}{(2\pi)^3}\, k^2r_d^2\frac{A^0_\text{w}(k)^2 \cos^2(k r_d)}{2 W_2(k)^2} e^{-2 k^2 \Sigma_\Lambda(r_d)}\,,\nonumber\\
    \simeq&L^3 \int \frac{d^3 k}{(2\pi)^3}\, k^2r_d^2\frac{A^0_\text{w}(k)^2 \cos^2(k r_d)}{2 W_2(k)^2}\left(1- 2 k^2 \Sigma_\Lambda(r_d)\right)\,,
    \label{eq:FMPSsoft}
\end{align}
where, in the derivative of $W_2$, we have kept only terms enhanced by $k^2 r_d^2$.
The second, negative term in parentheses represents the loss of information on the characteristic scale $r_d$ that is caused by modes with $q<\Lambda$.

This information is restored once we include the bispectrum--bispectrum block of the Fisher matrix, in particular its Gaussian piece (diagram a in Fig.~\ref{fig:FMB2}), where we focus on contributions with one soft internal momentum: 
\begin{align} 
F_{\beta,g}^{(3,3)}=&\frac{L^6}{2}\int_\Lambda \frac{d^3 p_1}{(2 \pi)^3}\frac{d^3 p_2}{(2 \pi)^3} \int^\Lambda\frac{d^3 q}{(2 \pi)^3}  \frac{(2 \pi)^3\delta_D(\bq+\bp_1+\bp_2)}{W_2(q)W_2(p_1)W_2(p_2)} \left(\frac{\partial W_3(q,p_1,p_2)}{\partial \beta}\right)^2\,.
\label{eq:FMbispsoft}
\end{align}
The soft limit of the bispectrum is fixed by the consistency relation \cite{Peloso:2013zw, Kehagias:2013yd},
\beq
L^3 W_3(k,|\bk+\bq|,q)\to \frac{k}{q}\mu \,b_1 P(q) \left(W_2(k)-W_2(|\bk+\bq|)\right) + O\big((q/k)^0\big)\,,
\label{eq:CRbisp}
\eeq
valid when $q/k \ll 1$, with $b_1$ the linear bias.

Restricting to the oscillatory (wiggly) part of the power spectrum, we obtain
\beq
W_{2,\text{w}}(k) - W_{2,\text{w}}(|\bk+\bq|)=  A_\text{w}(k)  \left[ \sin(k r_d)\left(1-\cos(q \mu r_d)\right)- \cos(k r_d) \sin(q \mu r_d)\right]\,.
\eeq
In the range $q\ll 2 \pi/r_d\ll k$, the right-hand side becomes 
$-A_\text{w}(k) \cos(k r_d) q \mu r_d + O(q^2 r_d^2) \simeq q \mu \; d W_{2,\text w}(k)/d k$, whereas for $2\pi/r_d\ll q\ll  k$ the oscillatory factors in $q\mu r_d$ average out and we can approximate this difference by $W_{2,\text w}(k)$ (see, for instance, \cite{Baldauf:2015xfa, Nishimichi:2017gdq}).  In summary, the soft limit of the bispectrum acquires a wiggly contribution sourced by short modes of the form
\begin{equation}
  L^3  W_{3,\text{w}}(k,|\bk+\bq|,q)  \to   \left\{\begin{array}{lc}
   -\mu^2  b_1 P(q)\,k\frac{d W_{2\text w}(k)}{dk}& (q\ll 2\pi/r_d\ll k)\\
   &\\
   \mu \frac{k}{q}  b_1 P(q)\, W_{2\text w}(k) &  (2\pi/r_d\ll q\ll  k)
   \end{array}\right.\,.
   \label{eq:softbisp}
\end{equation}
Using these relations, we can evaluate the contribution of soft modes $q$ to the bispectrum Fisher matrix. Modes in the interval $2 \pi r_d^{-1}\ll q\ll k$ contribute to Eq.~\eqref{eq:FMbispsoft} as
\beq
\frac{1}{2}\int_\Lambda \frac{d^3 k}{(2 \pi)^3} \, k^2r_d^2\frac{A^0_\text{w}(k)^2 \cos^2(k r_d)}{ W_2(k)^2}\frac{2}{3}\int_{2 \pi r_d^{-1}}^\Lambda \frac{d^3 q}{(2 \pi)^3}\frac{k^2}{q^2} \frac{b_1^2 P(q)^2}{W_2(q)}\,,
\eeq
where we have accounted for the two configurations $\bp_1=\bk$, $\bp_2=-\bk-\bq$ and $\bp_2=\bk$, $\bp_1=-\bk-\bq$, and we have approximated $W_2(|\bk+\bq|)\simeq W_2(k)$ in the denominator. When we add this term to the negative contribution in Eq.~\eqref{eq:FMPSsoft} for the same momentum configurations, we obtain
\beq
-\frac{L^3}{3}\int_\Lambda \frac{d^3 k}{(2 \pi)^3} k^2 r_d^2\frac{ A^0_\text{w}(k)^2 \cos^2(k r_d)}{W_2(k)^2}\int_{2 \pi r_d^{-1}}^\Lambda \frac{d^3 q}{(2\pi)^3}\frac{k^2}{q^2}P(q) \, \frac{P_\epsilon}{b_1^2 P(q)+P_\epsilon}\,,
\label{eq:FMpart}
\eeq
where for the soft $q$ modes we have assumed the model
\beq
L^3 W_2(q)= b_1^2 P(q)+P_\epsilon,
\eeq
with $P_\epsilon$ denoting the noise power spectrum. 
Equation~\eqref{eq:FMpart} describes the residual information loss due to very long modes, which is controlled by the noise amplitude.

The contribution of modes with wavenumbers $q\ll2\pi r_d^{-1}\ll k$ to Eqs.~\eqref{eq:FMPSsoft} and \eqref{eq:FMbispsoft} yields the following residual information loss:
\begin{align}
&-\frac{L^3}{10}\int_\Lambda \frac{d^3k}{(2\pi)^3} \frac{k^4 r_d^4 A^0_\text{w}(k)^2}{W_2(k)^2}\int^{2\pi r_d^{-1}}\frac{d^3q}{(2\pi)^3} P(q)\left(\cos^2(k r_d)-\frac{b_1^2 P(q)}{L^3 W_2(q)}\sin^2(k r_d)  \right)\,,
\nonumber \\
&\simeq -\frac{L^3}{20}\int_\Lambda \frac{d^3k}{(2\pi)^3} \frac{k^4 r_d^4 A^0_\text{w}(k)^2}{W_2(k)^2}\int^{2\pi r_d^{-1}}\frac{d^3q}{(2\pi)^3} P(q)\,\frac{P_\epsilon}{b_1^2P(q)+P_\epsilon}\,,
\end{align}
where, in the second line, we have replaced $\sin^2(k r_d)$ and $\cos^2(k r_d)$ by their averages. This is justified because $k r_d\gg 2\pi$. After performing this averaging, the information loss in this regime is again controlled by the noise amplitude.

Higher powers of the damping factor $k^2 \Sigma_\Lambda(r_d)$ are obtained by evaluating the Fisher matrix of higher-order correlators that contain only two hard modes, while the remaining modes are soft. In this case, Eq.~\eqref{eq:softbisp} generalizes to
\begin{align}
L^{3j}\,W_{2+j,\text{w}}&(\bk,-\bk-\bq_{1\cdots j},\bq_1,\cdots,q_{j_\text{t}},q_{j_\text{t}+1},\cdots,\bq_j) \nonumber\\
&\to (-1)^{j_\text{t}} \,\left(\Pi_{i=1}^{j_\text{t}} \mu_i^2 b_1 P(q_i)\right)\left(\Pi_{l={j_\text{t}+1}}^j \mu_l\frac{k}{q_l} b_1 P(q_l)\right)\;k^{j_\text{t}}\frac{d^{j_\text{t}} W_{2,\text{w}}(k)}{dk^{j_\text{t}}}\,,\nonumber \\&\qquad\qquad\qquad\qquad\qquad\qquad\qquad (q_i\ll 2\pi/r_d\ll k,\quad 2\pi/r_d\ll q_l\ll  k)\,.
\end{align}
We have thus shown that the FLP includes the information from the soft (squeezed) limits of all correlators, combining them to optimally recover the information on the scale $r_d$, given the noise. In the evaluation of the FLP, correlators are summed over all configurations however. Hence, this information recovery is only under control when the full prediction for the higher-order correlators considered is under control over all configurations.

Further, we should highlight that the sole consideration of $F_{\beta\beta}$ implies that all other parameters are kept fixed. In practice, in analysis using the FLP, one would also vary the other parameters that are relevant for predicting the correlators $W_n$. This includes in particular the bias parameters. Ref.~\cite{babic/etal:2026} argues that the broad-band contribution to the correlators could further enhance the information in the FLP over low-order summary statistics. It would be interesting to confirm this by extending the single-parameter Fisher calculation to the multi-parameter case. However, it might be difficult to obtain closed analytical results in this case.

\section{Reconstruction of the initial field}
\label{sect:reco}
We now show how the  formalism developed so far naturally yields, as a byproduct, the optimal reconstruction of the initial conditions of the field.
To achieve this, we adjust Eq.~\eqref{eq:fllikM} by adding a source term for the initial conditions,
\begin{align}
\hat {\cal L}[\delta_d;J_{\rm in}]=&\int {\cal D}\delta_{\rm M}\, {\cal D}\delta_{\rm in}\, {\cal D} \epsilon\; \tilde{\delta}_D\left[ \delta_d-\delta_{\rm M}\right]\,\tilde{\delta}_D\left[ \delta_{\rm M}-\bar\delta[\delta_{\rm in},\epsilon]\right]\,\,{\cal P}[\delta_{\rm in},\,\epsilon]\,e^{-i J_{\rm in} \cdot \delta_{\rm in}}\,,\nonumber\\
=&\int{\cal D} J_f \;e^{i J_f \cdot \delta_d}\, e^{ \hat W[J_f,J_{\rm in}]}\,,
\label{eq:L2} 
\end{align}
where
\beq
e^{ \hat W[J_f,J_{\rm in}]}\equiv \int {\cal D}\delta_{\rm M} \,{\cal D}\delta_{\rm in}  \hat P_M[\delta_{\rm M},\delta_{\rm in}] e^{-\left(i J_{\rm in} \cdot \delta_{\rm in}+ J_f \cdot \delta_{\rm M}\right)}\,,
\eeq
with
\beq \hat P_M[\delta_{\rm M},\delta_{\rm in}]\equiv \int{\cal D}\epsilon \,{\cal P}[\delta_{\rm in},\,\epsilon]\,\tilde{\delta}_D\left[ \delta_{\rm M}-\bar\delta[\delta_{\rm in},\epsilon]\right]\,.
\eeq
$\hat W[J_f,J_{\rm in}]$ is the generating functional of the connected mixed correlation functions, namely,
\begin{align}
&\left. i^{n+m}\left(\frac{\delta \;}{\delta\,J_f(\bn_1)}\cdots \frac{\delta \;}{\delta\,J_f(\bn_n)}\frac{\delta \;}{\delta\,J_{\rm in}(\bm_1)}\cdots \frac{\delta \;}{\delta\,J_{\rm in}(\bm_m)} \right) \hat W[J_f,J_{\rm in}]\right|_{J_{\rm in}=J_f=0}\nonumber\\
&\qquad \equiv W_{n,m}(\bn_1,\cdots,\bn_n,\bm_1,\cdots,\bm_m)\,,
\end{align}
where $W_{n,m}(\bp_1,\cdots,\bp_n,\bq_1,\cdots,\bq_m)$ is the connected correlator of $n$ final (that is, nonlinear) fields and $m$ initial (linear) ones; as before, we absorb a total momentum delta function into the definition. It then follows that, when $m=0$, we recover the correlation functions of the final field, namely, 
\beq
W_{n,0}=W_n\,.
\eeq
Notice that
\beq
\hat {\cal L}[\delta_d;J_{\rm in}=0]={\cal L}[\delta_d]\,,
\label{eq:l2l0}
\eeq
and $\hat W[J_f,J_{\rm in}=0]=W[J_f]$.
We define the {\em average} initial field for given data $\delta_d$ and source $J_{\rm in}$ as
\beq
\tilde \delta_{\rm in}[\delta_d, J_{\rm in}](\bn)\equiv i \frac{\delta \log \hat {\cal L}[\delta_d;J_{\rm in}]}{\delta J_{\rm in}(-\bn)}\,,
\label{eq:deltav}
\eeq
which is a functional of the data field $\delta_d$ and of the source term $J_{\rm in}$, and,
by a Legendre transform, we define the likelihood
\begin{equation}
L[\delta_d, \tilde \delta_{\rm in}] \equiv \hat {\cal L}[\delta_d;J_{\rm in}] \, e^{-i J_{\rm in}\cdot \tilde \delta_{\rm in}}\,.
\label{eq:defL}
\end{equation}
The reconstructed ``classical'' field $\delta^c_{\rm in}[\delta_d]$ is the field configuration that extremizes $\log L$ for given $\delta_d$,
\begin{equation}
\frac{\delta \log L[\delta_d, \delta^c_{\rm in}]}{\delta \delta^c_{\rm in}[\bm]}= 0\,.
\end{equation}
Since, from \eqref{eq:defL} we have,
\beq
\frac{\delta \log L[\delta_d, \tilde\delta_{\rm in}]}{\delta \tilde\delta_{\rm in}(\bm)}=-i J_{\rm in}(-\bm),
\eeq
 the  classical  field is given by
\beq
\delta^c_{\rm in}[\delta_d](\bn)\equiv \tilde \delta_{\rm in}[\delta_d, J_{\rm in}=0](\bn)\,,
\label{eq:deltaclass}
\eeq
and, moreover,
\beq
L[\delta_d,\delta^c_{\rm in}]={\cal L}[\delta_d]\,.
\eeq

From Eq.~\eqref{eq:L2} we have
\beq
\delta^c_{\rm in}[\delta_d](\bn)= {\cal L}[\delta_d]^{-1}\int {\cal D}J_f \,e^{i J_f\cdot \delta_d} e^{W[J_f]}\, \left.\frac{i \,\delta \hat W[J_f,J_{\rm in}]}{\delta J_{\rm in}(-\bn)}\right|_{J_{\rm in}=0}\,.
\label{eq:deltacnew}
\eeq
We now expand the functional inside the integral as
\begin{align}
  \left.\frac{i\,\delta \hat W[J_f,J_{\rm in}]}{\delta J_{\rm in}(-\bn)}\right|_{J_{\rm in}=0}=&\sum_{n=1}^\infty \frac{(-i)^{n}}{n!} \sum_{m_1,\cdots,m_n}^{m_{\rm max}}W_{n,1}(\bm_1,\cdots,\bm_n,\bn)J_f(\bm_1)\cdots J_f(\bm_n)\,,
\end{align}
Inserting this expression into Eq.~\eqref{eq:deltacnew}, and converting the factors of $J_f$ into functional derivatives with respect to $\delta_d$, we finally obtain the expression of the classical reconstructed field in terms of derivatives of the FLP,
\begin{align}
    \delta^c_{\rm in}[\delta_d](\bn)=&{\cal L}[\delta_d]^{-1}\sum_{n=1}^\infty \frac{(-1)^{n}}{n!}\sum_{m_1,\cdots,m_n}^{m_{\rm max}}W_{n,1}(\bm_1,\cdots,\bm_n,\bn)\frac{\delta}{\delta \delta_d(-\bm_1)}\cdots \frac{\delta}{\delta \delta_d(-\bm_n)}{\cal L}[\delta_d]\,.
\label{eq:deltacsmart}
\end{align}

At lowest order in derivatives ($n=1$) and using the Gaussian approximation of the FLP,  ${\cal L}[\delta_d]\simeq{\cal L}_g[\delta_d]$, we get
\begin{align}
\delta^{c,(0)}_{in}[\delta_d](\bn) =& \frac{W_{1,1}(n)}{W_{2}(n)}\delta_d(\bn)\,.
\label{eq:delta0in}
\end{align}
To build some intuition for this result, we assume a simple linear model with noise, with the following two-point correlators,
\[ L^3 W_{2}^{(0)}(n)=(A^2 b_1^2 P^0(n)+P_\epsilon(n)) D_{f}^2, \qquad  L^3W_{1,1}^{(0)}(n)=A^2 b_1^2 P^0(n) D_{in} D_f\,,
\]
where $D_{in}$ and $D_f$ are the linear and final growth factors, respectively.
Inserted into \eqref{eq:delta0in}, this gives
\begin{align}
\delta^{c,(0)}_{in}[\delta_d](\bn) &=\frac{D_{in}}{D_f}\left(1+\frac{P_\epsilon(n)}{A^2 b_1^2P^0(n)}\right)^{-1}\delta_d(\bn)=\frac{D_{in}}{D_f}\delta_{d,\epsilon}(\bn)  \,,
\end{align}
where we have defined the noise-filtered field
\beq
\delta_{d,\epsilon}(\bn)\equiv \left(1+\frac{P_\epsilon(n)}{A^2 b_1^2 P^0(n)} \right)^{-1} \delta_d(\bn)\,.
\label{eq:filterd}
\eeq
This is the expected Wiener-filter result (e.g., \cite{Kostic:2022vok}).
The prefactor tends to unity in the noiseless regime (that is, for large scales such that $P_\epsilon(n)/P^0(n)\ll 1$), and is damped on small scales. 

At next order, we include the contribution of the three-point correlators $W_3$ and $W_{2,1}$:
\begin{align}
\delta^{c,(1)}_{in}[\delta_d](\bn)=& -\frac{1}{2}\sum_{m_1,m_2}^{m_{\rm max}}\bigg[\frac{W^{(0)}_{1,1}(n)}{W^{(0)}_{2}(n)} W_{3}^{(1)}(m_1,m_2,n)
\nonumber\\
&\qquad\qquad\qquad\qquad\qquad\qquad -W_{2,1}^{(1)}(m_1,m_2,n)\bigg]\frac{\delta_d(\bm_1)}{W_{2}^{(0)}(m_1)}\frac{\delta_d(\bm_2)}{W_{2}^{(0)}(m_2)}\,,\nonumber\\
=&- \frac{D_{in}}{D_f}  \Bigg\{ \left(1+\frac{P_\epsilon(n)}{A^2 b_1^2 P^0(n)} \right)^{-1} \sum_{m_1,m_2}^{ m_{\rm max}}  \Bigg[K_2(\bm_1,\bm_2) 
\nonumber\\
&\qquad\qquad\qquad- P_\epsilon(n) \left(\frac{K_2(-\bn,\bm_1)}{A^2 b_1^2 P^0(m_2) } +\frac{K_2(-\bn,\bm_2)}{A^2 b_1^2 P^0(m_1) } \right) \Bigg] \delta_{d,\epsilon}(\bm_1) \delta_{d,\epsilon}(\bm_2)\Bigg\}\,.
\end{align}
We have used the tree-level expressions,
\begin{align}
L^6 W_{3}^{(1)}[\bm_1,\bm_2,\bm_3]=&\delta_{\bm_1+\bm_2+\bm_3}\,2\, D_f^4 A^4 b_1^2\,\bigg[K_2(\bm_1,\bm_2)P^0(m_1)P^0(m_2)\nonumber\\ &\qquad\;+K_2(\bm_2,\bm_3)P^0(m_2)P^0(m_3)+K_2(\bm_3,\bm_1)P^0(m_3)P^0(m_1)\bigg]\,,\\
L^6 W_{2,1}^{(1)}[\bm_1,\bm_2,\bm_3]= &\delta_{\bm_1+\bm_2+\bm_3}\,2 \, D_f^3 D_{in} A^4 b_1^2\bigg[K_2(\bm_2,\bm_3)P^0(m_2)P^0(m_3)\nonumber\\
&\qquad+K_2(\bm_3,\bm_1)P^0(m_3)P^0(m_1)\bigg]\,,
\end{align}
where the difference between the two expressions is due to the fact that in $W_{2,1}[\bp_1,\bp_2,\bp_3]$ the field carrying momentum $\bp_3$ is evaluated at the initial time. The contribution coming from its second-order correction is therefore suppressed by a factor $D_{in}/D_f$ with respect to the other two and is neglected. 
It is straightforward to check that, in the noiseless limit, this result coincides with that of the inverse PT of \cite{Cabass2024}. However, besides holding also in the noisy regime, this approach provides a more direct derivation of the connection to the correlators.

In PT, the next order will include contributions involving two nonlinearities, that is $F_3$ and $F_2 \times F_2$. It will include terms from the one-loop power spectrum, two tree-level bispectra, and one tree-level trispectrum.

\section{Summary and outlook}
\label{sect:concl}
In this paper we have developed a field-level posterior (FLP) for large-scale structure and other cosmological observables in a form that is fully non-perturbative with respect to the forward model, and that can therefore be applied equally well to perturbative calculations, simulation-based predictions, or more general effective descriptions. The only basic assumptions entering the construction are translational invariance, the existence of a probability distribution for the initial conditions and the noise, and the possibility of organizing the result in terms of connected correlators of the evolved field. Analytical progress can then be made in case of a Gaussian PDF of the initial conditions, the relevant case for cosmology.

Expanding the FLP around its Gaussian limit, we derive a general decomposition of the Fisher matrix into contributions labeled by the connected correlators $W_j$. This makes it possible to identify, term by term, which pieces of the full field-level information are retained by likelihoods for compressed observables and which are instead lost when the data are reduced to a finite set of summary statistics. In particular, we recovered the standard power spectrum Fisher matrix from the Gaussian part of the likelihoods for  the power spectrum  and the bispectrum, and clarified how cross-covariances among different summaries reconstruct part of the information that is absent in single-summary analyses.

A central message of our analysis is that the FLP provides a unified language to compare field-level inference with correlator-based likelihoods. From this viewpoint, the information contained in the FLP can be reorganized as the information carried by an infinite hierarchy of correlators, provided one keeps not only their individual likelihoods but also their full covariance and cross-covariance structure. Truncating this hierarchy at the level of the power spectrum, of the bispectrum, or of a finite joint data vector such as $P+B$, leads to a precise and quantifiable information loss, which we expressed in terms of the missing higher-order connected correlators.
An advantage of this expansion is that it allows one to apply the same truncation scheme as in standard correlator analyses (e.g., $P^{\rm 1-loop} + B^{\rm tree}$). In contrast, a field-level forward model is always truncated at a fixed perturbative order \cite{Cabass2024,Schmidt:2025iwa}; for instance, at second order it would contain the $P^{(22)}$ contribution to the 1-loop power spectrum but omit the $P^{(13)}$ one.
Our treatment also makes explicit the regime of validity of the expansion around the Gaussian limit. The relevant control parameters are set by combinations of higher-point correlators and powers of $W_2$, and are closely related to the quantities that govern the signal-to-noise of non-Gaussian observables. This provides a useful criterion for assessing when Gaussian likelihoods for summary statistics are adequate, when non-Gaussian covariance corrections are required, and when one should instead resort to the full field-level description.

We have illustrated these ideas in several concrete settings. For the BAO scale, the main lesson is not only that the FLP clarifies the relation between field-level and correlator-based extractions of the information, but also that it makes explicit how the field contains all the information needed for an optimal reconstruction of the BAO scale in the presence of noise. In this sense, the FLP identifies which terms in the expansion are required to saturate that optimal limit, and therefore provides a systematic way to understand what is retained or lost in approximate reconstruction schemes based on compressed observables. For reconstruction of the initial conditions, we showed that the classical initial field emerges naturally as a byproduct of the formalism itself. The resulting estimator is the optimal reconstruction of the initial field given the observed data and the noise, and it connects in the appropriate limit to the Wiener filter (when truncating at the power spectrum) as well as previous perturbative inverse methods (at higher order). These examples underline that the formalism is not restricted to abstract information-theory questions, but also provides a practical framework to analyze commonly used inference strategies within a single language.

There are several natural directions for future work. A first one is to extend the present formalism to redshift-space distortions, survey windows, and multi-tracer observables; we note that the formalism developed here directly applies to redshift-space distortions, by inserting predictions for redshift-space correlators for the $W_j$. A second is to investigate approximation schemes or resummations that remain under control beyond the regime where the expansion in connected correlators holds. A third is to exploit the flexibility of the FLP in situations where the forward model is provided by simulations, emulators, or machine-learning surrogates, so that the correlators entering the expansion can be measured or inferred without relying on a strict perturbative treatment. 

Another important direction concerns the treatment of noise. Recent work has emphasized that non-Gaussian noise must be handled carefully in order to recover unbiased field-level inference \cite{Akitsu:2025boy,Rubira:2025rqo}. In our approach, noise properties enter directly at the correlator level, through the $W_j$'s, rather than via explicit noise realizations drawn from an assumed probability distribution. This suggests that incorporating realistic noise and exploring its impact within the present framework may benefit from established EFTofLSS results for correlators. Finally, it would be interesting to explore in more detail the relation between the expansion parameters of the FLP and the signal-to-noise of higher-order observables, in order to sharpen the practical criterion for when field-level inference yields a substantial gain over compressed-likelihood approaches.

In this sense, the FLP is not only a formal generalization of correlator likelihoods, but also a tool to organize systematically the comparison between different inference strategies in large-scale structure, identify their domains of validity, and quantify the information that is gained or lost by compression. Future research should further specify how to carry out an efficient numerical implementation of the method introduced in this paper.

\section*{Acknowledgements}
\noindent
We thank the Galileo Galilei Institute and the organizers of the workshop ``New physics from galaxy clustering at GGI,'' where this work was initiated. We also thank Giorgia Biselli, Henrique Rubira, Marko Simonovi\'c, and Atsushi Taruya for discussions, and Guido D'Amico for providing useful feedback after a careful reading of the draft. 

\appendix
\section{Conventions}
\label{app:conventions}
In this appendix we collect the conventions for Fourier transforms and connected correlators used throughout the paper.

We work in a cubic box of side $L$, with periodic boundary conditions, and label Fourier modes by integer triplets
\beq
\bm=(m_x,m_y,m_z)
\eeq
with associated wavevector
\beq
\bk_\bm=k_f\,\bm\,,\qquad k_f\equiv \frac{2\pi}{L}\,.
\eeq
The fundamental field variable used in the paper is the \emph{dimensionless} Fourier-space density field,
\beq
\delta(\bm)\equiv \frac{1}{L^3}\int_{L^3} d^3x\; e^{-i \bk_\bm\cdot \bx}\,\delta(\bx)
\,.
\label{eq:appFT}
\eeq
On the discrete grid this is equivalent to the convention used in Eq.~\eqref{eq:FCadim}, namely dividing the Fourier transform by the box volume $L^3$ (or, equivalently on the lattice, by the total number of grid points). With this choice, both $\delta(\bx)$ and $\delta(\bm)$ are dimensionless. Moreover, since $\delta(\bx)$ is real,  Fourier space modes satisfy the constraint,
$$
\delta(-\bm)=\delta(\bm)^\ast\,.
$$

Translational invariance implies momentum conservation. Throughout the paper we absorb the corresponding Kronecker delta and the $L^3$ factor into the definition of the connected correlators and write
\beq
\langle \delta(\bm_1)\cdots \delta(\bm_j)\rangle_c
=\,W_j(\bm_1,\ldots,\bm_j)\,,
\label{eq:appWnDef}
\eeq
where $\bm_{1\cdots j}\equiv \bm_1+\cdots+\bm_j$. The quantities $W_j$ are therefore the reduced, dimensionless connected correlators that appear in the FLP expansion, in agreement with the convention adopted in the main text.

For the lowest orders, the relation with the standard connected correlators is
\begin{align}
\langle \delta(\bm_1)\delta(\bm_2)\rangle_c
&=W_2(\bm_1,\bm_2)
=\delta_{\bm_1+\bm_2}\,P(\bk_1)L^{-3}\,,\nonumber\\
\langle \delta(\bm_1)\delta(\bm_2)\delta(\bm_3)\rangle_c
&=W_3(\bm_1,\bm_2,\bm_3)
=\delta_{\bm_{123}}\,B(\bk_1,\bk_2,\bk_3)\,L^{-6}\,,\nonumber\\
\langle \delta(\bm_1)\cdots\delta(\bm_4)\rangle_c
&=W_4(\bm_1,\bm_2,\bm_3,\bm_4)
=\delta_{\bm_{1234}}\,T(\bk_1,\bk_2,\bk_3,\bk_4)\,L^{-9}\,,
\label{eq:WB}
\end{align}
and similarly for higher connected correlators. In other words, with our Fourier convention, the objects $W_2$, $W_3$, $W_4$, \ldots are the dimensionless counterparts of the usual power spectrum, bispectrum, trispectrum, \ldots.

More generally, if one indicates dimensional connected correlators of order $j$ as $C_j$
then our convention gives
\beq
W_j(\bm_1,\ldots,\bm_j)=\delta_{\bm_{1\cdots j}}\,C_j(\bk_1,\ldots,\bk_j)\,L^{3(1-j)}
\,.
\label{eq:appCnWn}
\eeq

Finally, for the two point function  we often abbreviate
\beq
W_2(\bm)\equiv W_2(\bm,-\bm)\,.
\eeq
 Analogous shorthand notations are used for bin-averaged bispectra, trispectra, and higher correlators when discussing compressed likelihoods.
 
\section{Properties of the \texorpdfstring{${\cal H}_j[\delta_d]$}{Hj[delta d]} functionals}
\label{app:Hj}
We define the  probabilist's Hermite functional polynomials as 
\begin{align}
{\cal H}_j[\delta_d](\bn_1,\cdots,\bn_j)
&\equiv (-1)^j\,{\cal L}_g[\delta_d] ^{-1} \frac{\delta^j}{\delta\delta_d(\bn_1)\cdots \delta\delta_d(\bn_j)}{\cal L}_g[\delta_d]\,,
\label{eq:Hdef}
\end{align}
where $\mathcal{L}_g[\delta_d]$ is defined in Eq.~\eqref{eq:L0gaussfull}.
Introducing an auxiliary field $\chi[\bn]$ and the generating functional 
\begin{align}
{\cal G}[\delta_d,\chi]&\equiv\sum_{j=0}^{\infty}\;\sum_{\bm_1,\cdots,\bm_j}\frac{1}{j!}\,{\cal H}_j(\bm_1,\cdots,\bm_j)\chi(\bm_1)\cdots \chi(\bm_j)\,,\nonumber\\
&=
\exp\left[-\frac{1}{2}\sum_{\bm}\frac{|\chi(\bm)|^2 - 2 \,\chi(\bm)\delta_d(-\bm)}{W_2(\bm)}\right]\,,
\end{align}
the generic ${\cal H}_j[\delta_d]$ functional can be computed as
\begin{align}
{\cal H}_j[\delta_d](\bn_1,\cdots,\bn_j)
&=\left. \frac{\delta^j}{\delta\chi(\bn_1)\cdots \delta\chi(\bn_j)}{\cal G}[\delta_d,\chi]\right|_{\chi=0}\,.
\end{align}
The averages with respect to the Gaussian FLP,  have the following properties
\begin{align}
&\langle {\cal H}_j[\delta_d](\bn_1,\cdots,\bn_j) \rangle_g=\delta_{j0}\,,\label{eq:avH}\\
&\langle {\cal H}_{j}[\delta_d](\bn_1,\cdots,\bn_{j}) {\cal H}_{j'}[\delta_d](\bn'_1,\cdots,\bn'_{j'}) \rangle_g=\delta_{j j'}\, \frac{\sum_{\sigma \in S_j}
\delta_{\bn_1+\bn'_{\sigma(1)}} \cdots \delta_{\bn_1+\bn'_{\sigma(j)}}}{ W_2(\bn_1) \cdots  W_2(\bn_j) }\,,\label{eq:avHH}
\end{align}
where $S_j$ is the permutation group of  $\{1,2,\cdots,j\}$.
Eq.~\eqref{eq:avH} can be shown directly from the definition \eqref{eq:Hdef} and the vanishing of the (functional) integral over derivatives of a functional vanishing at infinity,
\begin{align}
\langle {\cal H}_j[\delta_d](\bn_1,\cdots,\bn_j) \rangle_g=&(-1)^j\int {\cal D}\delta_d {\cal L}_g[\delta_d] {\cal L}_g[\delta_d] ^{-1} \frac{\delta^j}{\delta\delta_d(\bn_1)\cdots \delta\delta_d(\bn_j)}{\cal L}_g[\delta_d]\nonumber\\
=&(-1)^j\int {\cal D}\delta_d  \frac{\delta^j}{\delta\delta_d(\bn_1)\cdots \delta\delta_d(\bn_j)}{\cal L}_g[\delta_d]=\delta_{j0}\,.
\end{align}
As for the orthogonality condition, Eq.~\eqref{eq:avHH}, we can write
\begin{align}
&\langle {\cal H}_{j}[\delta_d](\bn_1,\cdots,\bn_j) {\cal H}_{j'}[\delta_d](\bn'_1,\cdots,\bn'_{j'}) \rangle_g\nonumber\\&=\int {\cal D}\delta_d \; \mathcal{L}_g[\delta_d]\,{\cal H}_{j}[\delta_d](\bn_1,\cdots,\bn_j){\cal H}_{j'}[\delta_d](\bn'_1,\cdots,\bn'_{j'})\,,\nonumber\\
&=(-1)^j\int {\cal D}\delta_d  \left(\frac{\delta^j}{\delta\delta_d(\bn_1)\cdots \delta\delta_d(\bn_j)}{\cal L}_g[\delta_d]\right){\cal H}_{j'}[\delta_d](\bn'_1,\cdots,\bn'_{j'})\,,\nonumber\\
&=\int {\cal D}\delta_d {\cal L}_g[\delta_d]  \frac{\delta^j}{\delta\delta_d(\bn_1)\cdots \delta\delta_d(\bn_j)}{\cal H}_{j'}[\delta_d](\bn'_1,\cdots,\bn'_{j'})\,,\nonumber\\
&=\int {\cal D}\delta_d {\cal L}_g[\delta_d]  \left.\frac{\delta^{j'+j}}{\delta \chi(\bn'_1)\cdots \delta \chi(\bn'_{j'})\delta\delta_d(\bn_1)\cdots \delta\delta_d(\bn_j)}{\cal G}_{j'}[\delta_d,\chi]\right|_{\chi=0}\,,\nonumber\\
&= \delta_{j j'}\, \frac{\sum_{\sigma \in S_j}
\delta_{\bn_1+\bn'_{\sigma(1)}} \cdots \delta_{\bn_j+\bn'_{\sigma(j)}}}{ W_2(\bn_1) \cdots  W_2(\bn_j) }\int {\cal D}\delta_d \;{\cal L}_g[\delta_d]\,, 
\end{align}
which equals the RHS of Eq.~\eqref{eq:avHH}, thanks to the normalization of ${\cal L}_g$.

Another important feature is the composition rule for two polynomials, which allows to express   the product ${\cal H}_{j}{\cal H}_{j'}$ as a sum of terms that are linear in the polynomials, involving only ${\cal H}_{j''}$  with $j''$ ranging from $|j-j'|$ to $j+j'$ in increments of two. The explicit form of the expansion can be derived by considering the product of generating functionals
\beq
{\cal G}[\delta_d,\chi_1] {\cal G}[\delta_d,\chi_2] = e^{-\frac{1}{2}\left[|\chi_1|^2+|\chi_2|^2- 2(\chi_1+\chi_2)\delta_d\right]}={\cal G}[\delta_d,\chi_1+\chi_2]e^{\chi_1 \chi_2}\,,
\eeq
where we have used a compact notation. 
Therefore,
\begin{align}
{\cal H}_{j}{\cal H}_{j'}=&\left.\frac{\delta^j}{\delta \chi_1^j}{\cal G}[\delta_d,\chi_1]\right|_{\chi_1=0} \left.\frac{\delta^{j'}}{\delta \chi_2^{j'}}{\cal G}[\delta_d,\chi_2]\right|_{\chi_2=0}\nonumber\\
=&\left.\frac{\delta^{j+j'}}{\delta \chi_1^j\delta \chi_2^{j'}}\left( {\cal G}[\delta_d,\chi_1+\chi_2]e^{\chi_1 \chi_2}\right) \right|_{\chi_1=\chi_2=0}\nonumber\\
=& {\cal H}_{j+j'}+\cdots\,.
\end{align}
As an example, in Eq.~\eqref{eq:FM22V2} we  encounter the product ${\cal H}_j {\cal H}_2(\bm_1,\bm_2)$, which, by applying the identity above, can be expressed as  
\begin{align}
    &{\cal H}_j(\bn_1,\cdots,\bn_j){\cal H}_2(\bm_1,\bm_2)= \;{\cal H}_{j+2}(\bn_1,\cdots,\bn_j,\bm_1,\bm_2)\nonumber\\
    &+\sum_{i=1}^j\frac{1}{ W_2(\bn_i)}\left(\delta_{\bn_i+\bm_1}{\cal H}_j(\bn_1,\cdots,\bn_{i-1},\bn_{i+1},\cdots,\bn_j,\bm_2)+(\bm_1\leftrightarrow \bm_2)\right)\nonumber\\
    &+\sum_{i=2}^j\sum_{l=1}^{i-1}\frac{1}{ W_2(\bn_i) W_2(\bn_l)}\left(\delta_{\bn_l+\bm_1}\delta_{\bn_i+\bm_2}+\delta_{\bn_l+\bm_2}\delta_{\bn_i+\bm_1}\right)\nonumber\\
    &\qquad\qquad\qquad\qquad\qquad\times {\cal H}_{j-2}(\bn_1,\cdots,\bn_{i-1},\bn_{i+1},\cdots,\bn_{l-1},\bn_{l+1},\cdots,\bn_j)\,.
\end{align}
Analogously, we get
\begin{align}
    &{\cal H}_j(\bn_1,\cdots,\bn_j){\cal H}_3(\bm_1,\bm_2,\bm_3)= \;{\cal H}_{j+3}(\bn_1,\cdots,\bn_j,\bm_1,\bm_2,\bm_3)\nonumber\\
    &+\sum_{i=1}^j\frac{1}{W_2(\bn_i)}\left(\delta_{\bn_i+\bm_1}{\cal H}_{j+1}(\bn_1,\cdots,\bn_{i-1},\bn_{i+1},\cdots,\bn_j,\bm_2, \bm_3)+ 2\;\text{cyclic perms.}\right)\nonumber\\
    &+\sum_{i=2}^j\sum_{l=1}^{i-1}\frac{1}{ W_2(\bn_i) W_2(\bn_l)}\big[\left(\delta_{\bn_l+\bm_1}\delta_{\bn_i+\bm_2}+\delta_{\bn_l+\bm_2}\delta_{\bn_i+\bm_1}\right)\nonumber\\
    &\qquad\qquad\qquad\times {\cal H}_{j-1}(\bn_1,\cdots,\bn_{i-1},\bn_{i+1},\cdots,\bn_{l-1},\bn_{l+1},\cdots,\bn_j,\bm_3)+ 2\;\text{cyclic perms.}\big]\nonumber\\
    &+\sum_{i=3}^j\sum_{l=2}^{i-1}\sum_{k=1}^{l-1}
 \sum_{\sigma \in S_3}\frac{\delta_{\bn_i+\bm_{\sigma(1)}}\delta_{\bn_l+\bm_{\sigma(2)}}\delta_{\bn_k+\bm_{\sigma(3)}}}{ W_2(\bn_i) W_2(\bn_l) W_2(\bn_k)}
 \nonumber\\
     &\qquad\qquad\qquad\times{\cal H}_{j-3}(\bn_1,\cdots,\bn_{i-1},\bn_{i+1},\cdots,\bn_{l-1},\bn_{l+1},\cdots,\bn_{k-1},\bn_{k+1},\cdots,\bn_j)\,.
\end{align}
\section{Computation of the Fisher matrix from the FLP to \texorpdfstring{$O(V_I^2)$}{O(VI2)}}
\label{app:FMcalc}
We expand the quantity inside the central parenthesis in Eq.~\eqref{eq:FMpar}, namely, 
\beq
\int {\cal D}\delta_d\,\frac{1}{{\cal L}[\delta_d]} \frac{\delta^i {\cal L}[\delta_d]}{\delta \delta_d^i}\frac{\delta^j {\cal L}[\delta_d]}{\delta \delta_d^j}\,,
\label{eq:FMcov}
\eeq
to second order in the non-Gaussian operator $V_I$. 
We express the $j$-th functional derivative of the FLP as
\begin{align}
    \frac{\delta^j {\cal L}}{\delta\delta_d(\bn_1)\cdots \delta\delta_d(\bn_j)} &=\frac{\delta^j}{\delta\delta_d(\bn_1)\cdots \delta\delta_d(\bn_j)}\left[\sum_{k=0}^\infty \frac{(V_I)^k}{k!} {\cal L}_g\right]\nonumber\\
    &=(-1)^j\,{\cal L}_g \sum_{k=0}^\infty\frac{1}{k!}\sum_{j_1,\cdots,j_k=3}^\infty \frac{1}{j_1!\cdots j_k!} \left[W_{j_1}\cdots W_{j_k}\right] \cdot {\cal H}_{j_1+\cdots +j_k+j}\,,
    \label{eq:jderL}
\end{align}
where
\begin{align}
\left[W_{j_1}\cdots W_{j_k}\right]& \cdot {\cal H}_{j_1+\cdots +j_k+j}(\{\bn\})\nonumber\\ 
&\equiv \sum_{\bn_1^1,\cdots,\bn^1_{j_1}}W_{j_1}(\bn^1_1,\cdots,\bn^1_{j_1})\cdots \sum_{\bn_1^k,\cdots,\bn_{j_k}^k}W_{j_k}(\bn^k_1,\cdots,\bn_{j_k}^k)\nonumber\\&\qquad\times{\cal H}_{j_1+\cdots +j_k}(\bn^1_1,\cdots,\bn^1_{j_1},\cdots,\bn^k_1,\cdots,\bn^k_{j_k},\bn_1,\cdots,\bn_j)\,.
\label{eq:Lderexp}
\end{align}
Using Eqs.~\eqref{eq:exp3} and \eqref{eq:Lderexp} in \eqref{eq:FMcov}, and keeping terms up to second order in $W_j$'s we get
\begin{align}
&\int \mathcal{D}\delta_d\,
\frac{1}{\mathcal{L}[\delta_d]}\,
\frac{\delta^i \mathcal{L}[\delta_d]}
{\delta \delta_d(\bm_1)\cdots \delta_d(\bm_i)}\,
\frac{\delta^j \mathcal{L}[\delta_d]}
{\delta \delta_d(\bn_1)\cdots \delta_d(\bn_j)}
\nonumber\\
&= \delta_{ij}\,
\left\langle
\mathcal{H}_i(\{\bm\})\,
\mathcal{H}_j(\{\bn\})
\right\rangle
\nonumber\\
&\quad
+ \sum_{j_1=3}^{\infty}
\frac{(-1)^{i+j}}{j_1!}\,
\, W_{j_1}\cdot
\Big[
\left\langle
\mathcal{H}_{j_1+i}(\{\bm\})\,
\mathcal{H}_j(\{\bn\})
\right\rangle
\nonumber\\
&\qquad
+ \left\langle
\mathcal{H}_{j_1+j}(\{\bn\})\,
\mathcal{H}_i(\{\bm\})
\right\rangle
- \left\langle
\mathcal{H}_{j_1}\,
\mathcal{H}_i(\{\bm\})\,
\mathcal{H}_j(\{\bn\})
\right\rangle
\Big]
\nonumber\\
&\quad
+ \sum_{j_1,j_2=3}^{\infty}
\frac{(-1)^{i+j}}{j_1!\,j_2!}\,
\big[ W_{j_1} W_{j_2} \big]\,
\Bigg[
\nonumber\\
&\qquad
\left\langle
\Big(
\mathcal{H}_{j_1+i}(\{\bm\})
- \mathcal{H}_{j_1}\mathcal{H}_i(\{\bm\})
\Big)
\Big(
\mathcal{H}_{j_2+j}(\{\bn\})
- \mathcal{H}_{j_2}\mathcal{H}_j(\{\bn\})
\Big)
\right\rangle
\nonumber\\
&\qquad
+ \frac{1}{2}
\left\langle
\mathcal{H}_{j_1+j_2+i}(\{\bm\})\,
\mathcal{H}_j(\{\bn\})
+ \mathcal{H}_i(\{\bm\})\,
\mathcal{H}_{j_1+j_2+j}(\{\bn\})
\right.
\nonumber\\
&\qquad\qquad
\left.
- \mathcal{H}_{j_1+j_2}\,
\mathcal{H}_i(\{\bm\})\,
\mathcal{H}_j(\{\bn\})
\right\rangle
\Bigg]\,,
\label{eq:FMcov2}
\end{align}
where $\mathcal{H}_i(\{\bm\})=\mathcal{H}_i(\{\bm\})[\delta_d]$ and we defined the average with respect to the Gaussian FLP as
\beq
\langle {\cal F}[\delta_d]\rangle\equiv \int {\cal D}\delta_d\, {\cal L}_g[\delta_d] {\cal F}[\delta_d]\,.
\eeq
Setting $i=j=2$ in Eq.~\eqref{eq:FMcov2}, and using the properties of the $\mathcal{H}_j$ functionals described in Appendix \ref{app:Hj}, we get the contribution to the Fisher matrix from the power spectrum,
\begin{align}
&F_{ab}^{(2,2)}
= \frac{1}{4}
\sum_{m_1,m_2}^{m_{\rm max}}
\frac{\partial W_2(\bm_1)}{\partial \alpha_a}\,
\frac{\partial W_2(\bm_2)}{\partial \alpha_b}
\Bigg\{
\nonumber\\
&\qquad
\left\langle
\mathcal{H}_2(\bm_1,-\bm_1)\,
\mathcal{H}_2(\bm_2,-\bm_2)
\right\rangle
\nonumber\\
&\qquad
- \frac{1}{4!}\, W_4\cdot
\left\langle
\mathcal{H}_4\,
\mathcal{H}_2(\bm_1,-\bm_1)\,
\mathcal{H}_2(\bm_2,-\bm_2)
\right\rangle
\nonumber\\
&\qquad
+ \sum_{j_1,j_2=3}^{\infty}
\frac{1}{j_1!\,j_2!}\,
\big[ W_{j_1} W_{j_2} \big]
\nonumber\\
&\qquad\qquad
\times
\left\langle
\Big(
\mathcal{H}_{j_1+2}(\bm_1,-\bm_1)
- \mathcal{H}_{j_1}\,
\mathcal{H}_2(\bm_1,-\bm_1)
\Big)
\right.
\nonumber\\
&\qquad\qquad\qquad
\left.
\Big(
\mathcal{H}_{j_2+2}(\bm_2,-\bm_2)
- \mathcal{H}_{j_2}\,
\mathcal{H}_2(\bm_2,-\bm_2)
\Big)
\right\rangle
\Bigg\} \,,\label{eq:FM22V2}
\end{align}
given explicitly in Eqs.~\eqref{eq:FMps}--\eqref{eq:FMPSFLP}.

\section{Fisher information for multiple summary statistics}
\label{app:fishermarg}

We have found that $(F_{P+B})_{\alpha\alpha}$ is different from $(F_{P})_{\alpha\alpha}$ even for a parameter $p_\alpha$ that only enters $P$, i.e. $W_2$. This initially surprising finding can be understood as follows.

Consider a data vector $\v{x}$ with $n_x$ elements whose expectation value $\bar{\v{x}}(\vec{p})$ depends on a $k_x$-dimensional parameter vector $\vec{p}$. The Fisher information on $\vec{p}$ for observable $\v{x}$ is
\begin{align}
  F_{p_\alpha p_\beta}(x) = \frac{\partial \bar{\v{x}}}{\partial p_\alpha} \cdot C_{(x)}^{-1} \cdot \frac{\partial \bar{\v{x}}}{\partial p_\beta}  \,,
\end{align}
where $C_{(x)}$ is the covariance of the data vector $\v{x}$. The error on $p_\alpha$ marginalized over the $p_{\beta\neq\alpha}$ is given by $[F(\v{x})]^{-1}_{\alpha\alpha}$.

Now consider adding an observable vector $\v{y}$ with length $n_y$ whose mean $\bar{\v{y}}(\vec q)$ depends on another, $k_y$-dimensional parameter set $q$, but not $p$. This is the case considered in the derivation in Sect.~\ref{sect:FLPFM}. 
Then, the joint derivative matrix is $k\times n$, where $k\equiv k_x+k_y$,
$n\equiv n_x+n_y$, and block-diagonal:
\be
D \equiv \frac{\partial(\v{x},\v{y})}{\partial (\vec p, \vec q)}
= \left(
\begin{array}{cc}
  \frac{\partial \bar{\v{x}}}{\partial \vec p} & 0 \\
0 &   \frac{\partial \bar{\v{y}}}{\partial \vec q}
\end{array}
\right)\,.
\ee
We will further assume that the parameters are non-degenerate, which implies that $D$ has full row-rank $k$. We can then write the full Fisher matrix as
\be
F(x\oplus y) = D C_{(x\oplus y)}^{-1} D^\top\,.
\ee
It is then easy to see that the $\vec p$-block of is not equal to $F(x)$,
since the $xx$-block of $C_{(x\oplus y)}^{-1}$ is not equal to $C_{(x)}^{-1}$ if $x$
and $y$ are covariant, i.e. if the $xy$-block of $C_{(x\oplus y)}$ is nonzero This is what was found in \refeq{FMPSFLP}.

The difference arises because $F(x\oplus y)_{\alpha\alpha}$ corresponds to
the information on $p_\alpha$ while keeping all parameters in $\vec p$ and $\vec q$
apart from $p_\alpha$ fixed. Even though $\v{y}$ by itself does not contain
information on $p_\alpha$, the Fisher information is effectively modified due
to the covariance of $\v{x}$ and $\v{y}$.

The proper comparison between the $x$-only and $x\oplus y$ cases is to
marginalize over the parameters $\vec q$, thus removing the information
that comes from $\v{y}$. This is accomplished by inverting the Fisher matrix.
We can do this by using $D^\top (D D^\top)^{-1}$ as the right pseudo-inverse of $D$. This yields
\ba
F^{-1}(x\oplus y) &= (D D^\top)^{-1} D C_{(x\oplus y)} D^\top (D D^\top)^{-1} 
\ea
where
\be
(D D^\top)_{\alpha\beta} = \frac{\partial(\v{x},\v{y})}{\partial (\vec p, \vec q)_\alpha}
\cdot \frac{\partial(\v{x},\v{y})}{\partial (\vec p, \vec q)_\beta}
\ee
and the contraction is done over the $n$-dimensional combined observable vector. Our assumption that $\v{x}$ and $\v{y}$ depend on disjoint parameter sets
implies that $D D^\top$ is block-diagonal. In this case, one can easily verify
that the $\vec{p}$-block of $F^{-1}(x\oplus y)$ is the same as $F^{-1}(x)$.

That is, once marginalizing over the parameters controlling the additional
data vector $\v{y}$ (e.g., the bispectrum), the information on the parameters
$\vec{p}$ is the same as when considering only the data vector $\v{x}$ (e.g., the power spectrum). It is worth highlighting that this is independent of the
precise model $\bar{\v{y}}(\vec q)$ assumed. One canonical choice for this
would be to set $\bar{\v{y}} = \vec q$, i.e. simply allow every element of the
bispectrum to be a free parameter. 

This equivalence no longer holds as soon as one or more parameters
control both $\bar{\v{x}}$ and $\bar{\v{y}}$, such that the matrix $D$ is
no longer block-diagonal. In cosmology, this is usually the case.

\section[Computation of the Fisher matrix components]{Computation of the Fisher matrix components $F_{ab}^{(3,3)}$ and $F_{ab}^{(2,3)}$ to $O(V_I^2)$}
We split the Fisher contribution associated with the bispectrum as
\begin{align} F_{ab}^{(3,3)}&=F_{ab,\text{g}}^{(3,3)}+F_{ab,\text{B{cov}}}^{(3,3)}+F_{ab,\text{FLP}}^{(3,3)}\,,
   \label{eq:bispV2}
\end{align}
where the Gaussian part is given by
\begin{align}
  F_{ab,\text{g}}^{(3,3)}=&   \frac{1}{(3!)^2}  \sum_{\{m_i\}}^{m_{\rm max}}3!  \frac{\left(\frac{\partial W_3(\bm_1,\bm_2,\bm_3) }{\partial \alpha_a} \right) \left(\frac{\partial W_3(\bm_1,\bm_2,\bm_3) }{\partial \alpha_b}\right)}{ W_2(\bm_1)W_2(\bm_2)W_2(\bm_3)}\,.
\end{align}
The part also present in the Fisher matrix for the bispectrum, from the expansion of the inverse covariance, is
\begin{align}
F_{ab,\text{B{cov}}}^{(3,3)}=&\frac{1}{(3!)^2}  \sum_{\{m_i\},\{n_j\}}^{m_{\rm max}}\frac{\partial W_3(\bm_1,\bm_2,\bm_3) }{\partial \alpha_a}  \frac{\partial W_3(\bn_1,\bn_2,\bn_3) }{\partial \alpha_b} \Bigg\{\nonumber\\
    &  -\frac{W_6(\bm_1,\bm_2,\bm_3,\bn_1,\bn_2,\bn_3)}{ W_2(\bm_1)\cdots W_2(\bn_3)}\nonumber\\
    &-9 \frac{\delta_{\bm_1+\bn_1}}{ W_2(\bm_1)} \frac{W_4(\bm_2,\bm_3,\bn_2,\bn_3)}{ W_2(\bm_2)W_2(\bm_3) W_2(\bn_2) W_2(\bn_3)}\nonumber\\
    &-\frac{9}{2} \frac{W_3(\bm_1,\bn_2,\bn_3)W_3(\bn_1,\bm_2,\bm_3)}{ W_2(\bm_1)\cdots W_2(\bn_3)}\nonumber\\
    &+\frac{1}{3!}\sum_{\{\bn'\}}\frac{W_6(\bm_1,\bm_2,\bm_3,\bn'_1,\bn'_2,\bn'_3 )W_6(-\bn'_1,-\bn'_2,-\bn'_3,\bn_1,\bn_2,\bn_3)}{ W_2(\bm_1)\cdots W_2(\bn_3)W_2(\bn'_1)W_2(\bn'_2) W_2(\bn'_3) }\nonumber\\
    &+18 \sum_{\bn'}\frac{W_4(\bm_1,\bn_2,\bn_3,\bn') W_4(\bn_1,\bm_2,\bm_3,-\bn')}{ W_2(\bm_1)\cdots W_2(\bn_3)W_2(\bn') }\Bigg\}\,.
\end{align}
Finally, the contributions intrinsic to the FLP are
\begin{align}
&F_{ab,\text{FLP}}^{(3,3)}= \frac{1}{(3!)^2}  \sum_{\{m_i\},\{n_j\}}^{m_{\rm max}}\frac{\partial W_3(\bm_1,\bm_2,\bm_3) }{\partial \alpha_a}  \frac{\partial W_3(\bn_1,\bn_2,\bn_3) }{\partial \alpha_b} \Bigg\{\nonumber\\
    &+ \sum_{j=5}^\infty \;\sum_{n'_1,\cdots, n'_{j-3}}^{m_{\rm max}}\frac{1}{(j-3)!}\Bigg[ \nonumber\\
&+(1-\delta_{j6})\frac{W_j(\bm_1,\bm_2,\bm_3,\bn'_1,\cdots, \bn'_{j-3})W_j(\bn_1,\bn_2,\bn_3,\bn'_1,\cdots, \bn'_{j-3})}{ W_2(\bm_1)\cdots W_2(\bn_3)W_2(\bn_1')\cdots W_2(\bn'_{j-3}) }\nonumber\\
    &\quad 18 \frac{W_j(\bm_1,\bn_2,\bn_3,\bn'_1,\cdots, \bn'_{j-3})W_j(\bn_1,\bm_2,\bm_3,\bn'_1,\cdots, \bn'_{j-3})}{ W_2(\bm_1)\cdots W_2(\bn_3)W_2(\bn_1')\cdots W_2(\bn'_{j-3}) }\Bigg]\nonumber\\
    &+\sum_{j=3}^\infty \;\sum_{n'_1,\cdots, n'_{j-1}}^{m_{\rm max}} \Bigg[\frac{18\, \delta_{\bm_2+\bn_2}\delta_{\bm_3+\bn_3}}{(j-1)!} \nonumber\\
    &\times \frac{W_j(\bm_1,\bn'_1,\cdots, \bn'_{j-1})W_j(\bn_1,\bn'_1,\cdots, \bn'_{j-1}) }{ W_2(\bm_1) W_2(\bm_2)W_2(\bm_3)W_2(\bn_1) W_2(\bn'_1) \cdots W_2(\bn'_{j-1})} \nonumber\\
    &+\frac{18\, \delta_{\bm_3+\bn_3}}{(j-1)!}\nonumber\\
    &\times \Bigg( \frac{W_j(\bm_1,\bn'_1,\cdots,\bn'_{j-1})W_{j+2}(\bn_1,\bn_2,\bm_2,\bn'_1,\cdots,\bn'_{j-1})  }{W_2(\bm_1)W_2(\bm_2)W_2(\bm_3)W_2(\bn_1)W_2(\bn_2)W_2(\bn'_1)\cdots W_2(\bn'_{j-1})} +(\{\bm\} \leftrightarrow \{\bn\})\Bigg)\nonumber\\
    &+ \frac{3}{(j-1)!} \Bigg(\frac{W_j(\bm_1,\bn'_1,\cdots,\bn'_{j-1})W_{j+4}(\bn_1,\bn_2,\bn_3,\bm_2,\bm_3,\bn'_1,\cdots,\bn'_{j-1})  }{ W_2(\bm_1)W_2(\bm_2)W_2(\bm_3)W_2(\bn_1)W_2(\bn_2)W_2(\bn_3) W_2(\bn'_1)\cdots W_2(\bn'_{j-1})}\nonumber\\
   &+(\{\bm\} \leftrightarrow \{\bn\}) \Bigg)\Bigg]
    \nonumber\\
    &+\sum_{j=3}^\infty\, \sum_{n'_1,\cdots, n'_{j-2}}^{m_{\rm max}} \Bigg[\frac{\delta_{\bm_3+\bn_3}}{(j-2)!} \Bigg(\nonumber\\
    &\frac{36 \;W_j(\bm_1,\bn_2,\bn'_1,\cdots, \bn'_{j-2})W_j(\bn_1,\bm_2,\bn'_1,\cdots, \bn'_{j-2}) }{ W_2(\bm_1) W_2(\bm_2)W_2(\bm_3)W_2(\bn_1)W_2(\bn_2) W_2(\bn'_1) \cdots W_2(\bn'_{j-2})}\nonumber\\
    & + \frac{9 \;W_j(\bm_1,\bm_2,\bn'_1,\cdots, \bn'_{j-2})W_j(\bn_1,\bn_2,\bn'_1,\cdots, \bn'_{j-2}) }{ W_2(\bm_1) W_2(\bm_2)W_2(\bm_3)W_2(\bn_1)W_2(\bn_2) W_2(\bn'_1) \cdots W_2(\bn'_{j-2})} \Bigg)\nonumber\\
    &+\frac{9}{(j-2)!}\Bigg( \frac{W_j(\bm_1,\bn_3,\bn'_1,\cdots,\bn'_{j-2})W_{j+2}(\bn_1,\bn_2,\bm_2,\bm_3,\bn'_1,\cdots,\bn'_{j-2})  }{W_2(\bm_1)W_2(\bm_2)W_2(\bm_3)W_2(\bn_1)W_2(\bn_2)W_2(\bn_3)W_2(\bn'_1)\cdots W_2(\bn'_{j-2})}\nonumber\\
   &+(\{\bm\} \leftrightarrow \{\bn\}) \Bigg)\nonumber\\
   & +\frac{3}{(j-2)!} \Bigg( \frac{W_j(\bm_1,\bm_2,\bn'_1,\cdots,\bn'_{j-2})W_{j+2}(\bn_1,\bn_2,\bn_3,\bm_3,\bn'_1,\cdots,\bn'_{j-2})  }{W_2(\bm_1)W_2(\bm_2)W_2(\bm_3)W_2(\bn_1)W_2(\bn_2)W_2(\bn_3)W_2(\bn'_1)\cdots W_2(\bn'_{j-2})}\nonumber\\
   &+(\{\bm\} \leftrightarrow \{\bn\}) \Bigg)\Bigg]\Bigg\}.
\end{align}

The off-diagonal part of the field-level Fisher matrix involving the power spectrum and the bispectrum is given by
\begin{align}
F_{ab}^{(2,3)}&=F_{ab,\text{(P+B){cov}}}^{(2,3)}+F_{ab,\text{FLP}}^{(2,3)}\,,
       \label{eq:PBVII}
\end{align}
where the contributions appearing in the Fisher matrix for the $P+B$ data vector are 
\begin{align}
  &F_{ab,\text{(P+B){cov}}}^{(2,3)}=   \frac{1}{2! 3!}\sum_{m, \{n_j\}}^{m_{\rm max}} \frac{\partial W_2 (m)}{\partial \alpha_a}\frac{\partial W_3 (\bn_1,\bn_2,\bn_3)}{\partial \alpha_b}\Bigg\{\nonumber\\
  &\qquad\qquad\qquad-6 \,\delta_{\bm+\bn_1}\frac{W_3(-\bm,\bn_2,\bn_3)}{ W_2(\bm)^2W_2(\bn_2)W_2(\bn_3)}\nonumber\\
  &\qquad\qquad\qquad-\frac{W_5(\bm,-\bm,\bn_1,\bn_2,\bn_3)}{ W_2(\bm)^2\, W_2(\bn_1)W_2(\bn_2)W_2(\bn_3)}\nonumber\\
  &\qquad\qquad\qquad+\frac{1}{2}\sum_{n'_1,n'_2}^{m_{\rm max}}\frac{ W_4(\bm,-\bm,\bn'_1,\bn'_2)W_5 (-\bn'_1,-\bn'_2,\bn_1,\bn_2,\bn_3)}{W_2(\bm)^2W_2(\bn_1)W_2(\bn_2)W_2(\bn_3)W_2(\bn'_1)W_2(\bn'_2)}\nonumber\\
  &\qquad\qquad\qquad+\frac{1}{3!}\sum_{n'_1,n'_2,n'_3}^{m_{\rm max}} \frac{ W_5(\bm,-\bm,\bn'_1,\bn'_2,\bn'_3)W_6 (-\bn'_1,-\bn'_2,-\bn'_3,\bn_1,\bn_2,\bn_3)}{W_2(\bm)^2W_2(\bn_1)W_2(\bn_2)W_2(\bn_3)W_2(\bn'_1)W_2(\bn'_2)W_2(\bn'_3)}\Bigg\}\,,
\end{align}
while the contribution appearing only at field level is
\begin{align}
F_{ab,\text{FLP}}^{(2,3)}=&    \frac{1}{2! 3!}\sum_{m, \{n_j\}}^{m_{\rm max}} \frac{\partial W_2 (m)}{\partial \alpha_a}\frac{\partial W_3 (\bn_1,\bn_2,\bn_3)}{\partial \alpha_b}\Bigg\{
\nonumber\\
&\!\!\!\!\!\!\!\!\!\!\!+\sum_{j=6}^\infty \frac{1}{(j-2)!}\sum_{n'_1,\cdots,n'_{j-2}}^{m_{\rm max}}\frac{ W_{j}(\bm,-\bm,\bn'_1,\cdots,\bn'_{j-2})W_{j+1} (-\bn'_1,\cdots,-\bn'_{j-2},\bn_1,\bn_2,\bn_3)}{W_2(\bm)^2W_2(\bn_1)W_2(\bn_2)W_2(\bn_3)W_2(\bn'_1)\cdots W_2(\bn'_{j-2})}
\nonumber\\
&\sum_{j=3}^\infty \Bigg[ \frac{6\, \delta_{\bn_1+\bm}}{(j-1)!}  \frac{W_j(\bm,\bn'_1,\cdots,\bn'_{j-1}) W_{j+1}(\bn_2,\bn_3,\bn'_1,\cdots,\bn'_{j-1})}{ W_2(\bm)^2 W_2(\bn_2)W_2(\bn_3)W_2(\bn'_1)\cdots W_2(\bn'_{j-1})} \nonumber\\
&+\sum_{n'_1,\cdots,n'_{j-2}}^{m_{\rm max}} \frac{3}{(j-2)!}\frac{W_j(\bn_2,\bn_3,\bn'_1,\cdots,\bn'_{j-2}) W_{j+1}(\bn_1,\bm,-\bm,\bn'_1,\cdots,\bn'_{j-2})}{ W_2(\bm)^2 W_2(\bn_1) W_2(\bn_2)W_2(\bn_3)W_2(\bn'_1)\cdots W_2(\bn'_{j-2})}\nonumber\\
&+\sum_{n'_1,\cdots,n'_{j-1}}^{m_{\rm max}}\frac{2}{(j-1)!}\frac{W_j(\bm,\bn'_1,\cdots,\bn'_{j-1}) W_{j+3}(\bn_1,\bn_2,\bn_3,-\bm,\bn'_1,\cdots,\bn'_{j-1})}{ W_2(\bm)^2 W_2(\bn_1) W_2(\bn_2)W_2(\bn_3)W_2(\bn'_1)\cdots W_2(\bn'_{j-1})}\nonumber\\
&+\sum_{n'_1,\cdots,n'_{j-1}}^{m_{\rm max}} \frac{3}{(j-1)!}\frac{W_j(\bn_1,\bn'_1,\cdots,\bn'_{j-1}) W_{j+3}(\bn_2,\bn_3,\bm,-\bm,\bn'_1,\cdots,\bn'_{j-1})}{ W_2(\bm)^2 W_2(\bn_1) W_2(\bn_2)W_2(\bn_3)W_2(\bn'_1)\cdots W_2(\bn'_{j-1})}\nonumber\\
&+\sum_{n'_1,\cdots,n'_{j-2}}^{m_{\rm max}}\frac{6}{(j-2)!} \frac{W_j(\bn_1,\bm,\bn'_1,\cdots,\bn'_{j-2}) W_{j+1}(\bn_2,\bn_3,-\bm,\bn'_1,\cdots,\bn'_{j-2})}{ W_2(\bm)^2 W_2(\bn_1) W_2(\bn_2)W_2(\bn_3)W_2(\bn'_1)\cdots W_2(\bn'_{j-2})}\nonumber\\
&+\sum_{n'_1,\cdots,n'_{j-1}}^{m_{\rm max}} \frac{12\, \delta_{\bn_1+\bm}}{(j-1)!}\frac{W_j(\bn_2,\bn'_1,\cdots,\bn'_{j-1}) W_{j+1}(\bn_3,\bm,\bn'_1,\cdots,\bn'_{j-1})}{W_2(\bm)^2 W_2(\bn_2)W_2(\bn_3)W_2(\bn'_1)\cdots W_2(\bn'_{j-1})}\Bigg]
\Bigg\}\,.
\end{align}
\section{On the Gaussian approximation to the power spectrum likelihood}
\label{app:corrLik}
In this appendix, we present an alternative derivation of the power spectrum likelihood that relies on the FLP expansion of correlators, and we emphasize the corrections to the standard Gaussian approximation given in Eq.~\eqref{eq:LPsgauss}.
\subsection{Gaussian contribution}
We begin with the Gaussian likelihood, Eq.~\eqref{eq:L0gaussfull}, which, inserted in \eqref{eq:ZOJ}, gives
\begin{align}
Z_{P,g}[J]=e^{W_{P,g}[J]}&= \int {\cal D}\delta_{\rm M}\, e^{-i J\cdot \hat P[\delta_{\rm M}]}\,e^{-\frac{1}{2} \delta_d W_2^{-1}\delta_d} \det[2 \pi W_2]^{-1/2}\,,
\nonumber\\
&=\int \left( \Pi_\bm^{m_{\rm max}}\frac{d \delta_{\rm M}(\bm)}{ \sqrt{2 \pi \,W_2(\bm)}}\right) e^{-\frac{1}{2 }\sum_\bm^{m_{\rm max}} \left(\frac{1}{W_2(\bm)}+2 i \frac{J(\bm)}{ N(\bm)}\right)|\delta_{\rm M}(\bm)|^2}\,,
\label{eq:ZOg}
\end{align}
where the product  and the sum run over all modes $\bm$, and we have introduced the functions $J(\bm)$ and $N(\bm)$, which evaluate to $J_l$ and $N_l$, respectively, for $\bm \in V_l$\,.
Redefining the field variable,
\beq
\tilde \delta_{\rm M}(\bm) \equiv D(\bm)^{1/2}\,\delta_{\rm M}(\bm) =\left(1+ 2 i \frac{J(\bm) W_2(\bm)}{ N(\bm)}\right)^{1/2} \, \delta_{\rm M}(\bm)\,,
\label{eq:Dmdef}
\eeq
the Gaussian integrals in \eqref{eq:ZOg} give
\begin{align}
 e^{W_{P,g}[J]}&=  \Pi_\bm^{m_{\rm max}}  \left(1+ 2 i \frac{J(\bm) W_2(\bm)}{ N(\bm)}\right)^{-1/2}\,,\nonumber\\
&= \Pi_{l=1}^{N_{\rm bin}}\, \Pi_{\bn \in V_l}  \left(1+ 2 i \frac{J_l W_2(n)}{ N_l}\right)^{-1/2}\,\nonumber\\
&\simeq \Pi_{l=1}^{N_{\rm bin}}\, \Pi_{\bn \in V_l}  \left(1+ 2 i \frac{J_l W_2^l}{ N_l}\right)^{-1/2}\,\nonumber\\
&=\exp\left[{-\frac{1}{2}\sum_{l=1}^{N_{\rm bin}} N_l \log\left(1+2 i\frac{J_l W_2^l}{ N_l}\right)}\right]\,,
\end{align}
where, for $\bn \in V_l$ we have approximated each $W_2(\bn)$ with its bin averaged value. In the large $N_l$ regime, the logarithm can be expanded, to give,
\begin{align}
W_{P,g}[J]=\sum_{l=1}^{N_{\rm bin}} \sum_{n=1}^{\infty} \frac{(-i)^n}{n}\left(\frac{2}{N_l}\right)^{n-1}\left(J_l W_2^l\right)^n\,,
\label{eq:WOgexp}
\end{align}
which is also an expansion in the sources $J_l$. Truncating at second order, and inserting in Eq.~\eqref{eq:LOZ}, we get the Gaussian likelihood for the power spectrum,
\begin{align}
    {\cal L}_{P,g}[\hat P_d]&=\Pi_{l=1}^{N_{\rm bin}} \int \frac{dJ_l}{2\pi}\;e^{i J_l (\hat P_d^l L^{-3}-W_2^l)-\frac{(J_l W_2^l)^2}{ N_l}}\,,
    \nonumber\\
    &=\Pi_{l=1}^{N_{\rm bin}} \;\frac{e^{-N_l \frac{(\hat P_d^l L^{-3}-W_2^l)^2}{4(W_2^l)^2}}}{\sqrt{4\pi L^6\frac{(W_2^l)^2}{N_l}}}\,,
\end{align}
where the covariance is given at leading order by
\beq
\Sigma^{lm}\simeq \Sigma^{lm}_g= \frac{2 }{N_l} (W_2^l)^2 \,\delta_{lm}\,. 
\eeq

From this likelihood we can compute the Fisher matrix. Allowing all the  $W_2^l$ to depend on the parameters, we can check that the maximum of the likelihood corresponds to $\hat P_d^l=L^3 W_2^l+O(1/N_l)$, and the second derivatives computed on these values give
\beq
F^{(2,2)}_{PS;ab}=\left.\frac{\partial^2}{\partial \alpha_a\partial \alpha_b}\left(-\log {\cal L}_{P,g}\right)\right|_{\hat P_d=L^3 W_2}= \sum_{l=1}^{N_{\rm bin}}\frac{\partial W_2^l}{\partial \alpha_a} \frac{\partial W_2^l}{\partial \alpha_b}\left(2\frac{ (W_2^l)^2}{N_l}\right)^{-1}\left(1+O\left(\frac{1}{N_l}\right)\right)\,,
\eeq
where $O(1/N_l)$ corrections 
include terms coming from cubic and  higher orders in  \eqref{eq:WOgexp}, like
\beq
e^{W_{P,g}[J]}=e^{- i J_l W_2^l-\frac{(J_l W_2^l)^2}{ N_l}}\left(1+ i \frac{4}{3}\frac{(J_lW_2^l)^3}{ N_l^2} +O\left(\frac{1}{N_l^3}\right)\right)\,.
\eeq
\subsection{First-order non-Gaussian contribution}
\label{subs:NGPS1}
The first non-Gaussian correction to the field-level likelihood, from Eq.~\eqref{eq:LL0}, has the form
\beq
V_I\left[-i \frac{\delta}{\delta \delta_{\rm M}}\right]{\cal L}_g[\delta_{\rm M}]\,,
\eeq
which, inserted in \eqref{eq:ZOJ}, gives the  non-Gaussian correction $\Delta Z_P[J]$.
The first contribution, which is linear in the three point function $W_3$, is zero because it involves three derivatives of the Gaussian likelihood and thus leads to a Gaussian integral over an odd number of fields. We are therefore lead to compute the contribution linear in the four point function, 
\begin{align}
\Delta Z_{P,W_4}[J]&=\frac{1}{4!} \sum_{\bm_1,\cdots,\bm_4}^{m_{\rm max}}W_4(\bm_1,\cdots,\bm_4)\int {\cal D}\delta_{\rm M}\;e^{-i J\cdot \hat P[\delta_{\rm M}]}\frac{\delta^4}{\delta \delta_{\rm M}(\bm_1)\cdots \delta \delta_{\rm M}(\bm_4)}{\cal L}_g[\delta_{\rm M}]\,.
\label{eq:DZO4}
\end{align}
The integral to be performed can be written as (see  Eq.~\eqref{eq:ZOg})
\begin{align}
 &\int {\cal D}\delta_{\rm M}\;e^{-i J\cdot \hat P[\delta_{\rm M}]} {\cal L}_g[\delta_{\rm M}]\;{\cal H}_4[\delta_{\rm M}]\,
    \nonumber\\
    =&\int {\cal D}\delta_{\rm M}\;  e^{-\frac{1}{2 L^3}\sum_\bm^{m_{\rm max}} \frac{|\delta_{\rm M}(\bm)|^2}{ W_2(\bm)} D(m)}\;\;{\cal H}_4[\delta_{\rm M}]\,,\nonumber\\
    =&\int {\cal D}\left(\tilde\delta_{\rm M}/D^{1/2}\right)\;  e^{-\frac{1}{2 L^3}\sum_\bm^{m_{\rm max}} \frac{|\tilde \delta_{\rm M}(\bm)|^2}{ W_2(\bm)} }\;\;{\cal H}_4[\tilde\delta_{\rm M}/D^{1/2}]\,,
\end{align}
which is a Gaussian integral on the variable $\tilde \delta_{\rm M}$.
We get,
 \begin{align}
     &\Delta Z_{P,W_4}[J]=\nonumber\\
     &\;\; =e^{W_P,g[J]} \frac{1}{4!}\sum_{\bm,\bn}^{m_{\rm max}} W_4(\bm,-\bm,\bn,-\bn)\frac{3}{ W_2(\bm) W_2(\bn)}\left(\frac{1}{D(\bm)D(\bn)}-\frac{1}{D(\bm)}-\frac{1}{D(\bn)}+1\right)\,,\nonumber\\
     &\;\;\simeq e^{W_O,g[J]}\frac{1}{8}\sum_{l,m}^{N_{\rm bin}}\frac{ W_4^{l,m}}{ W_2^lW_2^m}N_l N_m\left(\frac{1}{D_l D_m}-\frac{1}{D_l}-\frac{1}{D_m}+1\right)\,,\nonumber\\
     &\;\;= e^{W_O,g[J]}\left[-\frac{1}{2} \sum_{l,m}^{N_{\rm bin}}W_4^{l,m}J_l J_m+O\left(\frac{1}{N_{l,m}}\right)\right]\,,
     \label{eq:DZO4b}
 \end{align}
 where we have defined the bin averaged trispectrum as 
 \begin{equation}
 W_4^{l,m} \equiv \frac{1}{N_l N_m}\sum_{\bm \in V_l}\sum_{\bn \in V_m}W_4(\bm,-\bm,\bn,-\bn)\,,
 \end{equation}
 and, again, approximated the $W_2(\bn)$ with their bin-averaged values.

 Summing Eq.~\eqref{eq:DZO4b} and $Z_{P,g}[J]$, we obtain, from  \eqref{eq:LOZ} after performing the integral in $J_i$,
 \begin{align}
 {\cal L}_P[\hat P_d]&\simeq  {\cal L}_{P,g}[\hat P_d]\Bigg[1-\frac{1}{4}\sum_{l,m=1}^{N_{\rm bin}}\frac{W_4^{l,m}}{ W_2^l W_2^m}\left(\delta_{lm} N_l-N_l N_m\frac{(\hat P_d^l L^{-3}-W_2^l)(\hat P_d^m L^{-3}-W_2^m)}{2 W_2^l W_2^m}\right)\Bigg]\,,
 \label{eq:LOng}
 \end{align}
 and the Fisher matrix elements are then
 \begin{align}
&F^{(2,2)}_{PS;ab}=\left.\frac{\partial^2}{\partial \alpha_a\partial \alpha_b}\left(-\log {\cal L}_{P}\right)\right|_{O_d=W_2}\,,\nonumber\\
&\simeq \sum_{l,m=1}^{N_{\rm bin}}\frac{\partial W_2^l}{\partial \alpha_a} \frac{\partial W_2^m}{\partial \alpha_b}\Bigg[\frac{N_l }{2(W_2^l)^2}\delta_{lm}-\frac{1}{4} N_l N_m\frac{W_4^{l,m}}{ (W_2^l)^2(W_2^m)^2}\Bigg]\left(1+O\left(\frac{1}{N_{l,m}}\right)\right)\,,
\label{eq:FMW2field2}
\end{align}
where, the $O(W_4)$ term is the binned version of the one obtained at the   field-level, given by the first line in Eq.~\eqref{eq:FM22NG}. Taking  the derivative of the trispectrum with respect to the parameters $\alpha_{a,b}$  gives contributions suppressed by  $O(1/N_{l,m})$.

The next contribution, still at linear order in $V_I$, comes from the connected six-point function $W_6$,
\begin{align}
\Delta Z_{P,W_6}[J]&=\frac{1}{6!} \sum_{\bm_1,\cdots,\bm_6}^{m_{\rm max}}W_6(\bm_1,\cdots,\bm_6)\int {\cal D}\delta_{\rm M}\;e^{-i J\cdot \hat P[\delta_{\rm M}]}\frac{\delta^6}{\delta \delta_{\rm M}(\bm_1)\cdots \delta \delta_{\rm M}(\bm_6)}{\cal L}_g[\delta_{\rm M}]\,.
\label{eq:DZO6}
\end{align}
Proceeding as for $\Delta Z_{P,W_4}[J]$, we get 
\begin{equation}
\Delta Z_{P,W_6}[J]= e^{W_P,g[J]}\left[\frac{i}{3!} \sum_{lmn}^{N_{\rm bin}}W_6^{l,m,n}J_lJ_m J_n+O\left(\frac{1}{N_{l,m,n}}\right)\right]\,,
\end{equation}
where
\begin{equation}
 W_6^{l,m,n} \equiv \frac{1}{N_l N_m N_n}\sum_{\bm \in V_l}\sum_{\bn \in V_m}\sum_{\bl \in V_n}W_6(\bm,-\bm,\bn,-\bn,\bl,-\bl)\,,
 \label{eq:W6def}
 \end{equation}
 and, for a generic $2n$-point connected correlator, 
 \begin{equation}
\Delta Z_{P,W_{2n}}[J]= e^{W_P,g[J]}\left[\frac{(-i)^n}{n!} \sum_{i_1,\cdots, i_n}^{N_{\rm bin}}W_{2n}^{i_1,\cdots,i_n}J_{i_1}\cdots J_{i_n}+O\left(\frac{1}{N_{i_1,\cdots, i_n}}\right)\right]\,,
\label{eq:DZng}
\end{equation}
where the bin-averaged correlators have been defined analogously to $W_6^{i,j,k}$ in  \eqref{eq:W6def}.
Including all the non-Gaussian contribution at linear order in $V_I$, and at leading order in $1/N_i$, then gives the power spectrum likelihood, from
\eqref{eq:LOZ},
\begin{align}
 {\cal L}_P&[\hat P_d]\nonumber\\
 &\simeq  {\cal L}_{P,g}[\hat P_d]\Bigg[1+\sum_{n=2}^{\infty}\frac{1}{n!}\sum_{i_1,\cdots,i_n=1}^{N_{\rm bin}}\frac{W_{2n}^{i_1,\cdots,i_n}}{ W_2^{i_1} \cdots W_2^{i_n}}\left( \frac{(\hat P_d^{i_1} L^{-3}-W_2^{i_1})}{2 W_2^{i_1}/N_{i_1}}\cdots\frac{(\hat P_d^{i_n} L^{-3}-W_2^{i_n})}{2 W_2^{i_n}/N_{i_n}}\right) \Bigg]\,.
 \label{eq:LOngVI}
 \end{align}
 Notice, however, that the contributions for $n>2$ do not affect the Fisher matrix, which is still given by \eqref{eq:FMW2field2}.

\subsection{Higher-order non-Gaussian contributions}
Inserting into \eqref{eq:ZOJ} the second order non-Gaussian likelihood,
\beq
\frac{1}{2} \left(V_I\left[-i\frac{\delta}{\delta \delta_{\rm M}}\right]\right)^2{\cal L}_g[\delta_{\rm M}]\,,
\eeq
 we get contributions from bispectrum squared,
\begin{align}
    &\Delta Z_{P,W_3^2}[J]=\frac{1}{2}\left(\frac{-1}{3!}\right)^2\sum_{\bn_1\,\cdots,\bn_3}\sum_{\bn'_1\,\cdots,\bn'_3}\delta_{\bn_{1,\cdots, 3}}\delta_{\bn'_{1,\cdots, 3}}W_3(\bn_1,\cdots,\bn_3)W_3(\bn'_1,\cdots,\bn'_3)\nonumber\\
    &\quad\times\int {\cal D} \delta_{\rm M} e^{-i J\cdot \hat P[\delta_{\rm M}]}\frac{\delta^3}{\delta\delta_{\rm M}(\bn_1)\cdots \delta\delta_{\rm M}(\bn_3)}\frac{\delta^3}{\delta\delta_{\rm M}(\bn'_1)\cdots \delta\delta_{\rm M}(\bn'_3)}{\cal L}_g[\delta_{\rm M}]\,,
    \label{eq:W32}
\end{align}
and trispectrum squared,
\begin{align}
    &\Delta Z_{P,W_4^2}[J]=\frac{1}{2}\left(\frac{1}{4!}\right)^2\sum_{\bn_1\,\cdots,\bn_4}\sum_{\bn'_1\,\cdots,\bn'_4}\delta_{\bn_{1,\cdots, 4}}\delta_{\bn'_{1,\cdots, 4}}W_4(\bn_1,\cdots,\bn_4)W_4(\bn'_1,\cdots,\bn'_4)\nonumber\\
    &\quad\times\int {\cal D} \delta_{\rm M} e^{-i J\cdot \hat P[\delta_{\rm M}]}\frac{\delta^4}{\delta\delta_{\rm M}(\bn_1)\cdots \delta\delta_{\rm M}(\bn_4)}\frac{\delta^4}{\delta\delta_{\rm M}(\bn'_1)\cdots \delta\delta_{\rm M}(\bn'_4)}{\cal L}_g[\delta_{\rm M}]\,.
\end{align}
We are interested in extracting the leading contributions in the large $N_i$ limit.
To do so, we bring the functional derivatives on the source term in the integrals, leading to, in \eqref{eq:W32},
\begin{align}
    &\int \left( \Pi_\bm^{m_{\rm max}}\frac{d \delta_{\rm M}(\bm)}{ \sqrt{2 \pi \,W_2(\bm)}}\right) e^{-\frac{1}{2 }\sum_\bm^{m_{\rm max}} \frac{|\delta_{\rm M}(\bm)|^2}{W_2(\bm)}}\nonumber\\
    &\times \frac{\delta^3}{\delta\delta_{\rm M}(\bn_1)\cdots \delta\delta_{\rm M}(\bn_3)}\frac{\delta^3}{\delta\delta_{\rm M}(\bn'_1)\cdots \delta\delta_{\rm M}(\bn'_3)}\left(e^{-i\sum_\bm \frac{J(\bm)}{ N(\bm)}|\delta_{\rm M}(\bm)|^2}\right)\,,\nonumber\\
    &=\int \left( \Pi_\bm^{m_{\rm max}}\frac{d \delta_{\rm M}(\bm)}{ \sqrt{2 \pi \,W_2(\bm)}}\right)e^{-\frac{1}{2 }\sum_\bm^{m_{\rm max}} \frac{|\delta_{\rm M}(\bm)|^2}{W_2(\bm)}}e^{-i\sum_\bm^{m_{\rm max}} \frac{J(\bm)}{ N(\bm)}|\delta_{\rm M}(\bm)|^2}\nonumber\\
    &\times \frac{\delta^3}{\delta\delta_{\rm M}(\bn_1)\cdots \delta\delta_{\rm M}(\bn_3)}\frac{\delta^3}{\delta\delta_{\rm M}(\bn'_1)\cdots \delta\delta_{\rm M}(\bn'_3)}\left[ \frac{1}{3!}  \left(-i\sum_\bm^{m_{\rm max}} \frac{J(\bm)}{ N(\bm)}|\delta_{\rm M}(\bm)|^2\right)^3\right]+\cdots\,,\nonumber\\
    &= i 3!\,e^{W_{P,g}[J]}   \nonumber\\
    &\qquad\times\sum_{\bm,\bm',\bm''}^{m_{\rm max}} \frac{J(\bm)J(\bm')J(\bm'')}{N(\bm)N(\bm')N(\bm'')}\delta_{\bm_1+\bm}\delta_{\bm_2+\bm'}\delta_{\bm_3+\bm''}\delta_{\bm'_1-\bm}\delta_{\bm'_2-\bm'}\delta_{\bm'_3-\bm''}+\cdots\,,
\end{align}
where the dots indicate terms giving contributions subleading in $1/N(\bm)$.
Inserting in \eqref{eq:W32} we get
\begin{align}
 &\Delta Z_{P,W_3^2}[J]= \frac{i\, e^{W_{P,g}[J]}}{12}    \sum_{\bm,\bm'}^{m_{\rm max}} \frac{J(\bm)J(\bm')J(-\bm-\bm')}{N(\bm)N(\bm')N(-\bm-\bm')} \,\left(W_3(\bm,\bm',-\bm-\bm')\right)^2 +\cdots\,,
\end{align}
which is $O(1/N(\bm))$, since the sum runs only over two independent wavevectors, $\bm$, $\bm'$.

Proceeding in a similar way, we obtain the contribution quadratic in the trispectrum as,
\begin{align}
    &\Delta Z_{P,W_4^2}[J]=e^{W_{P,g}[J]}\Bigg\{\frac{1}{8}\left(\sum_{\bm,\bm'}^{m_{\rm max}} W_4(\bm,-\bm,\bm',-\bm')\frac{J(\bm)J(\bm')}{N(\bm)N(\bm')}\right)^2\nonumber\\
    &\qquad+\sum_{\bm,\bm'\bm''}^{m_{\rm max}} \Bigg[W_4(\bm,-\bm,\bm'',-\bm'')W_4(-\bm'',\bm'',\bm',-\bm')\frac{J(\bm)J(\bm')J(\bm'')J(-\bm'')}{N(\bm)N(\bm')N(\bm'')N(-\bm'')}\nonumber\\
    &\qquad+\frac{1}{3} \left(W_4(\bm,\bm',\bm'',-\bm-\bm'-\bm'')\right)^2 \frac{J(\bm)J(\bm')J(\bm'')J(-\bm-\bm'-\bm'')}{N(\bm)N(\bm')N(\bm'')N(-\bm-\bm'-\bm'')}\Bigg]\Bigg\}\,\nonumber\\
    & \qquad\qquad\quad=e^{W_{P,g}[J]} \frac{1}{8}\left(\sum_{j,k}^{N_{\rm bin}}W_4^{j,k}J_j J_k\right)^2+O\left(\frac{1}{N_i}\right)\,,
\end{align}
which, unlike $\Delta Z_{P,W_3^2}[J]$, contains an  unsuppressed term. Summing it  to \eqref{eq:DZng} we get, 
\begin{align}
Z_P[J]\simeq& e^{W_P,g[J]}\left[1 -\frac{1}{2}\sum_{j,k}^{N_{\rm bin}}W_4^{j,k}J_j J_k  + \frac{1}{2} \left(-\frac{1}{2}\sum_{j,k}^{N_{\rm bin}}W_4^{j,k}J_j J_k\right)^2  \right.\,\nonumber\\
&\qquad\qquad+\left.\sum_{n=3}^{\infty}\frac{(-i)^n}{n!} \sum_{i_1,\cdots, i_n}^{N_{\rm bin}}W_{2n}^{i_1,\cdots,i_n} J_{i_1}\cdots J_{i_n} +O\left(\frac{1}{N_i}\right)\right]\,.
\end{align}
Inspecting the structure of unsuppressed terms at higher orders, one  verifies that the $W_4$ contributions indeed exponentiate,
\beq
Z_P[J] \simeq e^{W_P,g[J]}\left[e^{-\frac{1}{2}\sum_{j,k}^{N_{\rm bin}}W_4^{j,k}J_j J_k}+\sum_{n=3}^{\infty}\frac{(-i)^n}{n!} \sum_{i_1,\cdots, i_n}^{N_{\rm bin}}W_{2n}^{i_1,\cdots,i_n}J_{i_1}\cdots J_{i_n}\right]\,.
\label{eq:Zres}
\eeq
Taking the Fourier transform, Eq.~\eqref{eq:LOZ},
the terms that involve powers of $W_4$ sum up to the full inverse covariance of the power spectrum, i.e. the inverse of Eq.~\eqref{eq:Sigma}, while terms involving $W_{\geq 6}$ lead to non-Gaussian corrections to the power spectrum likelihood:
\begin{align}
{\cal L}_P[\hat P_d]\simeq& 
\frac{1}{\sqrt{(2 \pi)^{N_{\rm bin}} \det(L^6\Sigma)}}\,\exp\!\left[-\frac{1}{2}\left(\hat P_d L^{-3}-W_2\right)\cdot \Sigma^{-1}\cdot \left(\hat P_d L^{-3}-W_2\right)\right]\,\nonumber\\
+&{\cal L}_{P,g}[\hat P_d]\sum_{n=3}^{\infty}\frac{1}{n!}\sum_{i_1,\cdots,i_n=1}^{N_{\rm bin}}\frac{W_{2n}^{i_1,\cdots,i_n}}{ W_2^{i_1} \cdots W_2^{i_n}}\left( \frac{(P_d^{i_1}L^{-3}-W_2^{i_1})}{2 W_2^{i_1}/N_{i_1}}\cdots\frac{(P_d^{i_n}L^{-3}-W_2^{i_n})}{2 W_2^{i_n}/N_{i_n}}\right)\,,
\label{eq:PSg}
\end{align}
with $\Sigma^{jk}$ the power spectrum covariance. The first line matches the Gaussian covariance form, Eq.~\eqref{eq:LPsgauss}, derived in Sect.~\ref{subsect:PSlik}, and thus reproduces the same Fisher matrix, namely Eq.~\eqref{eq:FPS}.

If, however, one is interested in the full likelihood for the power spectrum and not only in the Fisher matrix, the second line of Eq.~\eqref{eq:PSg}, which does not affect the Fisher matrix, must also be taken into account. Notice that, for example, the term quadratic in $W_4$ appearing in the first line is of higher order than the term linear in $W_6$ in the second line, when adopting a perturbation-theory power counting in the power spectrum.

\subsection{Likelihood for the bispectrum}
\label{sect:FMbisp}
The estimator for the bispectrum is
\beq
B^{t}[\delta]=\frac{1}{L^3 N_T^{t}}\sum_{\bm_1 \in V_i}\sum_{\bm_2 \in V_j}\sum_{\bm_3 \in V_k}\delta_{\bm_{123}}\,\delta(\bm_1) \delta(\bm_2) \delta(\bm_3)\,,
\eeq
where $t$ labels the set of closed triangles with side lengths in the momentum bins $V_i$, $V_j$, and $V_k$, and $N_T^{t}$ is the number of triangles in the bin $t$.
Following Eq.~\eqref{eq:likst}, we can define a likelihood for the bispectrum,
\beq
{\cal L}_B[B_d]=\int {\cal D}J \,e^{i J\cdot B_d}\, Z_B[J]\,,
\label{eq:likBS}
\eeq
with
\beq
 Z_B[J]=e^{W_B[J]}=\int{\cal D}\delta \,e^{-i J\cdot B[\delta]}\,{\cal L}[\delta]\,,
 \label{eq:WBdef}
\eeq
and
\begin{align}
    J\cdot B&\equiv \frac{1}{L^6}\sum_{t}^{\text{b.t.}}\, J_{t}\, B^{t}\,,\nonumber\\
    &=\frac{1}{L^9}\sum_{\bm_1 \in V_i}\sum_{\bm_2 \in V_j}\sum_{\bm_3 \in V_k}\delta_{\bm_{123}}\frac{J(\bm_1,\bm_2,\bm_3)}{N(\bm_1,\bm_2,\bm_3)}\,\delta(\bm_1) \delta(\bm_2) \delta(\bm_3)\,,
\end{align}
where the first sum is taken over all binned triangles.
We define
\beq
J(\bm_1,\bm_2,\bm_3)=J_{t}\,,\quad N(\bm_1,\bm_2,\bm_3)=N_T^{t}\,,\qquad \text{for}\;
\bm_1 \in V_i\,,\bm_2 \in V_j\,,\bm_3 \in V_k\,,
\eeq
and
\beq
{\cal D}J\equiv \Pi_{t}^{\text b.t.}\,\frac{d J_{t}}{2 \pi}\,.
\eeq
We then expand $W_B[J]$ around $J=0$,
\begin{align}
W_B[J]=&W_B[0]+\sum_{t}^{\text{b.t.}}\left.\frac{\partial W_B[J]}{\partial J_{t}}\right|_{J=0} J_{t}\nonumber\\
&+\frac{1}{2}\sum_{t,u}^{\text{b.t.}}\left.\frac{\partial^2 W_B[J]}{\partial J_{t} \partial J_{u}}\right|_{J=0} J_{t} \,J_{u}+ O(J^3)\,.
\end{align}
From Eq.~\eqref{eq:WBdef} we get,
\beq
W_B[0]=0\,,\qquad \left.\frac{\partial W_B[J]}{\partial J_{t}}\right|_{J=0}=-\frac{i}{L^6}W_3^{t}\,,
\eeq
where $W_3^{t}$ is the binned bispectrum of the model,
\beq
W_3^{t}\equiv \frac{1}{N_T^{t}}\sum_{\bm_1 \in V_i}\sum_{\bm_2 \in V_j}\sum_{\bm_3 \in V_k}\delta_{\bm_{123}}\,W_3(\bm_1,\bm_2,\bm_3)\,.
\eeq
Moreover,
\begin{align}
    \left.\frac{\partial^2 W_B[J]}{\partial J_{t} \partial J_{u}}\right|_{J=0}&=-\frac{1}{L^{12}}
\left[\langle B^{t}B^{u} \rangle -\langle  B^{t} \rangle\langle  B^{u} \rangle \right]\,,\nonumber\\
&\equiv - \Sigma_B^{t,u}\,,
\end{align}
is the bispectrum covariance. It can be decomposed into four contributions,
\beq
\Sigma_B^{t,u}=\delta_{t u}\Sigma_{g}^{t}+\Sigma_{W_3^2}^{t,u}+\Sigma_{W_2 W_4}^{t,u}+\Sigma_{W_6}^{t,u}\,,
\eeq
where the first `Gaussian' term includes contributions that are cubic in the power spectrum, the second involves terms quadratic in the bispectrum, and the third consists of products of the trispectrum with the power spectrum.
We therefore obtain
\begin{align}
    e^{W_B[J]}=&\,e^{-i J\cdot W_3-\frac{1}{2}J\cdot \Sigma_{W_2^3}\cdot J}\left[1-\frac{1}{2}J\cdot\left(\Sigma_{W_3^2}+\Sigma_{W_2 W_4}+\Sigma_{W_6}\right)\cdot J + O(J^3)\right]\,,
\end{align}
where, in analogy with the treatment of the power spectrum, only the first of the three covariance contributions has been exponentiated.
This gives the bispectrum likelihood via Eq.~\eqref{eq:likBS},
\begin{align}
    {\cal L}_B[B_d]=&\frac{1}{\sqrt{(2 \pi)^{N_{\rm bt}}\det (\Sigma_{g})}}
    \exp\!\left[-\frac{1}{2}\sum_{t}^{N_{\rm bt}}\frac{(B_d^{t}-W_3^{t})^2}{\Sigma_{g}^{t}}\right]\nonumber\\
    &\quad\times\Bigg(1-\frac{1}{2}\sum_{t,u}^{N_{\rm bt}}
    \frac{\Sigma_{W_3^2}^{t,u}+\Sigma_{W_2 W_4}^{t,u}+\Sigma_{W_6}^{t,u}}{\Sigma_{g}^{t}\Sigma_{g}^{u}}\nonumber\\
    &\qquad\qquad\times(B_d^{t}-W_3^{t})(B_d^{u}-W_3^{u})
    +O\left((B_d-W_3)\right)^3\Bigg)\,.
\end{align}
Computing the Fisher matrix from the Gaussian part, we recover the field-level result, Eq.~\eqref{eq:F33}.
This is the likelihood-level counterpart of the comparison between Eqs.~\eqref{eq:Fc33} and \eqref{eq:F33}. In the present notation, the Gaussian bispectrum likelihood leads to the same Fisher information once the sum over triangle bins $t$ and their covariance $\Sigma_g^t$ are identified with the standard bispectrum data vector and covariance.

\bibliographystyle{JHEP2015}
\bibliography{main}
\label{lastpage}
\end{document}